\documentclass[12pt,preprint]{aastex}

\usepackage{graphicx}
\usepackage{natbib}
\usepackage{color}
\usepackage{array}
\usepackage{rotating}
\usepackage{amsmath}
\usepackage{amssymb}
\usepackage{lineno}
\usepackage{appendix}

\usepackage{caption}
\usepackage{subcaption}

\newif\ifarxiv

\arxivtrue
\ifarxiv
\else
\linenumbers
\fi

\shorttitle{xxx}
\shortauthors{Wang and Wordsworth}

\begin{document}


\title{Extremely long convergence times in a 3D GCM simulation of the sub-Neptune Gliese 1214b}


\author{
Huize Wang\\
\vspace{0.1in}
\normalsize{School of Engineering and Applied Sciences, Geological Museum, 26 Oxford Street, Cambridge, MA 02138, USA}\\
Robin Wordsworth\\
\vspace{0.1in}
\normalsize{School of Engineering and Applied Sciences, Department of Earth and Planetary Sciences, Geological Museum, 26 Oxford Street, Cambridge, MA 02138, USA}\\
}

\begin{abstract}

We present gray gas general circulation model (GCM) simulations of the tidally locked mini-Neptune GJ 1214b. On timescales of 1,000-10,000 Earth days, our results are comparable to previous studies of the same planet, in the sense that they all exhibit two off-equatorial eastward jets. Over much longer integration times (50,000-250,000 Earth days) we find a significantly different circulation and observational features. The zonal-mean flow transitions from two off-equatorial jets to a single wide equatorial jet that has higher velocity and extends deeper. The hot spot location also shifts eastward over the integration time. Our results imply a convergence time far longer than the typical integration time used in previous studies. We demonstrate that this long convergence time is related to the long radiative timescale of the deep atmosphere and can be understood through a series of simple arguments. Our results indicate that particular attention must be paid to model convergence time in exoplanet GCM simulations, and that other results on the circulation of tidally locked exoplanets with thick atmospheres may need to be revisited. 

\end{abstract}


%\keywords{astrobiology---planet-star interactions---planets and satellites: atmospheres---techniques: spectroscopic}

\maketitle

\section{Introduction}

During the last few decades, more than 3700 planets orbiting other stars have been discovered and confirmed. The majority of these planets were discovered via the transit method. Many of the planets discovered by the transit method are close to their host star and are expected to be tidally locked, implying that only one side of the planet receives stellar radiation from the host star. This type of forcing provides the conditions for the development of rich and interesting atmospheric dynamics without any direct analog in solar System \citep{seager2010exoplanet,Showman2013}. 

Observing the atmospheric dynamics of exoplanets is difficult, but substantial progress has been made over the last few years. Currently, only a few observational features can be linked to atmospheric dynamics. The transit phase curve can be inverted to produce a longitudinal temperature distribution \citep[e.g., ][]{Knutson2007,demory2016map}, which can provide information on dynamical features such as zonal jets. The ``featureless'' transit spectra of the warm sub-Neptune/super-Earth\footnote{Previously, exoplanets such as GJ 1214b have often been referred to as ``super-Earths.'' Here we use the alternative term ``sub-Neptune'' for this planet and others of up to around 10 Earth masses that possess a thick atmosphere. }  GJ 1214b \citep{Charbonneau2009} and the Neptune-mass exoplanet GJ 436b \citep{knutson2014featureless,Kreidberg2014} hint at the existence of high and dense clouds, which require strong vertical updrafts in the upper atmosphere. \cite{crossfield2017trends} summarized the relationship between the amplitude of the spectral features of six sub-Neptunes and their planetary parameters. Super-Earths and sub-Neptunes are particularly interesting because they represent a class of planets that are not in our solar system. This makes them valuable targets for future observations and theoretical study.

Exoplanet atmospheric dynamics can be modeled in various ways. Three-dimensional general circulation models (GCMs) are widely used because of their ability to generate three-dimensional dynamical fields, model various physical processes, and simulate observational signals. For tidally locked planets, phase curves generated by GCMs have been used to predict observations \citep[e.g., ][]{zhang2017effects} and compare with observational results \citep[e.g., ][]{zellem20144}. Most of the GCMs are based on solving primitive equations, while \cite{mayne2019limits} pointed out that some assumptions of primitive equations break down in the regime of a sub-Neptune's atmosphere, and the strength of superrotation weakens when they solve the full Navier-Stokes equations. In this study, we focus on the GCMs that solve primitive equations to enable intercomparison with previous studies.

Multiple previous studies have used GCMs to study the dynamical regimes of strongly irradiated exoplanets with thick gas envelopes \citep[e.g., ][]{cho2003changing,Showman2009,showman2012doppler,heng2011atmospheric,rauscher2012general,komacek2016atmospheric,zhang2017effects,mayne2017results}. The atmospheric dynamics of tidally locked sub-Neptunes have also been investigated with GCMs. In particular, several previous studies have investigated the circulation of GJ 1214b \citep{Menou2012,zalucha2013investigation,Kataria2014,charnay2015mixing,Charnay2015cloud,drummond2018effect,mayne2019limits}. 

For exoplanets that have a significant gas envelope and are tidally locked to their host stars, superrotation and jet formation are discovered in many GCM simulations. These phenomena are believed to be linked to the shape of the thermal phase curve and shift of the hot spot \citep[e.g., ][]{Knutson2007}. Many theories have been developed for superrotation, jet formation, and atmospheric dynamics on tidally locked exoplanets, and their observational implications have also been explored \citep[e.g., ][]{Showman2011,tsai2014three,zhang2017effects,hammond2018wave}. Because there are still relatively few direct observational constraints on exoplanet circulation, these theories are anchored by comparison with the GCM results.

A major challenge in GCM simulations of a gas planet is that the deep atmosphere has a long equilibrium timescale, and model convergence is not well understood. Nonetheless, the standard approach in previous studies has been to initialize the atmosphere with an equilibrium vertical temperature profile that is horizontally homogeneous, and run the 3D GCMs for a few thousand Earth days (unless otherwise specified, all ``days'' in this paper are Earth days). Some previous studies have found that the choice of boundary condition and artificial drag can lead to differences in model results \citep[e.g., ][]{liu2013atmospheric,cho2015sensitivity,carone2019equatorial}. However, the dependence of flow structure on the equilibrium timescale of the deep atmosphere has only been investigated by a few previous studies. \cite{mayne2017results} studied the evolution of the deep atmosphere of the hot Jupiter HD 209458b, and found a long evolutionary timescale. \cite{carone2019equatorial} studied the dynamical feedback between the deep atmosphere and observable atmosphere by prescribing a Newtonian relaxation scheme in the deep atmosphere of their hot Jupiter simulations. They discussed how reducing the radiative time scale of the deep atmosphere modifies the flow structure. 

In the solar System, several previous studies have investigated processes that could be driving the jet structure at various depths on Jupiter and Saturn \citep[e.g., ][]{galperin2004ubiquitous,heimpel2005simulation, kaspi2007formation,read2007dynamics,,schneider2009formation,liu2010mechanisms,young2019simulating}. Most notably, recent gravity data from NASA's Juno mission have indicated that the jets observed in the surface weather layer of Jupiter extend thousands of meters deep into the atmosphere, probably to the depth at which magnetic dissipation becomes effective \citep{kaspi2018jupiter}. This suggests that similar dynamical connections between the observable weather layer and the deep atmosphere may exist on exoplanets. \cite{young2019simulating} ran GCM simulations of Jupiter for 130,000-150,000 Earth days to allow the deep region of the 18 bar atmosphere to come into equilibrium, and studied the dynamical properties of the Jovian atmosphere.

In this study, we present a suite of gray gas GCM simulations that we have performed using the Generic LMDZ GCM to investigate this issue. To allow intercomparison with a range of previous studies, our simulations use the same planetary parameters as those of the sub-Neptune GJ 1214b. Here we integrate our model for a much longer time than was done in previous work, with the aim of investigating the  convergence time. 

In Section~\ref{sec:method}, we describe the GCM and simulation setup. In Section~\ref{sec:results}, we show the model results, with a focus on the long timescale required for convergence. We also discuss how important dynamical features change with time, and consider the observational implications. In Section~\ref{sec:discussion}, we discuss the wider implications of our results, and give suggestions for future work. In Section~\ref{sec:conclusion} we present our summary and conclusions.

% section introduction (end)

\section{Method} % (fold)
\label{sec:method}

We used the generic LMDZ GCM, which has been developed specifically for modeling exoplanets and paleoclimate. It has previously been used to study the present and past climates of Earth, Mars, Venus, Titan, and exoplanets \citep{Wordsworth2011,Wordsworth2013,Forget2013,Leconte2013b,charnay2014titan}. The dynamical core \citep{Hourdin2006} solves the primitive equations using a finite difference method on an Arakawa C grid. The scheme is constructed to conserve enstrophy and total angular momentum, and scale-selective hyperdiffusion with a characteristic timescale of 16,000 s is used in the horizontal plane for stability \citep{Forget1999}. In this paper, we used a spatial resolution of 64 $\times$ 48 $\times$ 45 in longitude, latitude and altitude. The vertical layers are equally spaced in log-pressure, where the highest pressure is 80 bar (8 MPa) and the top level pressure is 1.3 Pa. The dynamical time step is 90 s, and the radiative time step is 450 s. For comparison, we also performed two sets of experiments where the highest pressure was 5 bar and 10 bar respectively. Here we mainly focus on the 80 bar experiment, which demonstrates the long equilibrium timescale most effectively. In Section~\ref{sec:results}, we discuss the effects of our choice of bottom boundary on the equilibrium timescale.

We focus on the planet GJ 1214b, which is a warm sub-Neptune \citep[planetary radius $R_P=2.68R_{\oplus}$, mass $M_P=6.55M_{\oplus}$; ][]{Charbonneau2009} orbiting an M dwarf. The planet's low density implies a thick atmosphere, and its relatively featureless transit spectrum suggests the presence of high and thick clouds \citep{bean2011optical,Kreidberg2014}. Here we assume that GJ 1214b has a circular orbit and is tidally locked to its host star, and that the stellar flux at the substellar point is 23,600 $\text{W}~\text{m}^{-2}$. We assume an internal energy flux of 0.73 $\text{W}~\text{m}^{-2}$. This value corresponds to an intrinsic temperature of 60 K, as suggested by \cite{rogers2010three}, and matches the estimation in \cite{thorngren2019intrinsic}. 

We used a two-stream gray gas radiative transfer scheme, with a shortwave mass absorption coefficient $\kappa_{\text{sw}}$ and a longwave mass absorption coefficient $\kappa_{\text{lw}}$. We chose $\kappa_{\text{sw}}=8\times10^{-5} ~ \text{m}^2~\text{kg}^{-1}$ and $\kappa_{\text{lw}}=2\times10^{-3} ~ \text{m}^2~\text{kg}^{-1}$ to match the globally averaged temperature profile with that of a 1D correlated-$k$ radiative-convective model \citep{miller2010nature,Menou2012} corresponding to an atmosphere of solar composition. Analytically, the 1D gray gas solution is (following \cite{Pierrehumbert2011BOOK})

\begin{equation}
T(p) = \left[\frac{\text{ASR}}{2\sigma} \left(1+\chi+(\chi^{-1}-\chi) \exp(- \frac{\kappa^I}{g \overline{\cos\theta}} \frac{p}{\chi} )\right)\right]^{0.25}
	\label{eq:graysolution}
\end{equation}

\begin{figure}[h]
    \centering
    \includegraphics[width=0.5\linewidth]{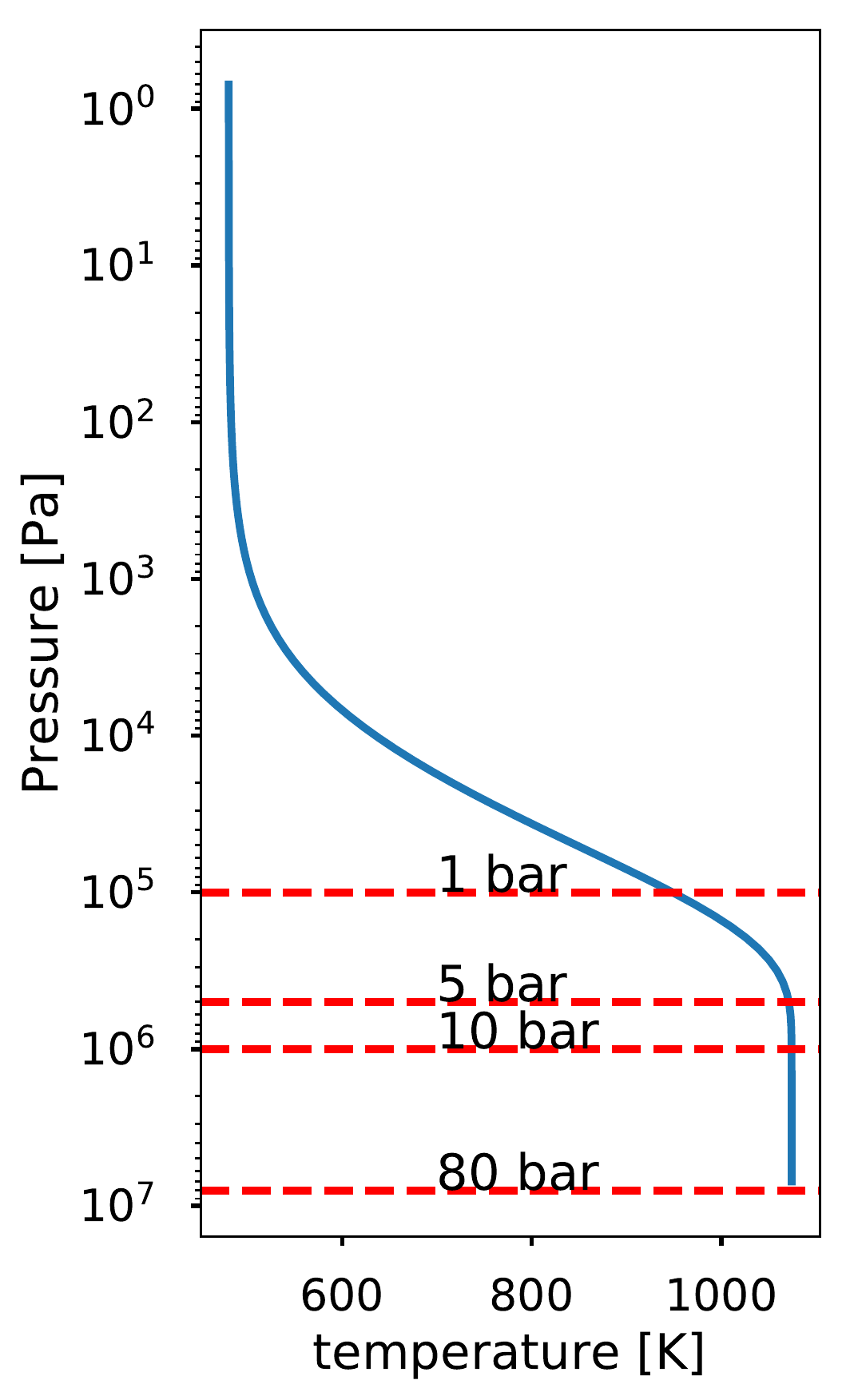}
    \caption{Analytic solution of temperature for a 1D gray gas model. The gray mass absorption coefficients are chosen to match the temperature profile of an atmosphere of solar composition calculated by \cite{miller2010nature}. When the pressure is greater than 100 bar, the $T$-$P$ profile gradually transitions to an adiabatic profile (Figure 1 of \cite{miller2010nature}).}
    \label{fig:graygas}
\end{figure}

Here $\chi = \kappa_{\text{lw}} / \kappa_{\text{sw}}$ and $\overline{\cos\theta}$ is the mean stellar zenith angle, which is assumed to be the same as the mean infrared propagation angle. Figure~\ref{fig:graygas} is the analytical solution to the 1D gray gas model, with $\overline{\cos\theta} = 1$. For the planetary parameters of GJ 1214b, this set of opacity values corresponds to $\tau_{\text{sw}}=1$ at $1.1$ bar pressure, and $\tau_{\text{lw}}=1$ at $44.7$ mbar, assuming an optical depth that increases moving downwards into the atmosphere.

In the upper levels of the atmosphere, a sponge layer is applied to reduce spurious reflections of vertically propagating waves. The sponge layer operates as a linear drag on the eddy components of the velocity fields, and does not change the zonal-mean velocities, as described in \cite{Forget1999}. The sponge layer term $- \frac{(A - \overline{A})}{\tau_\text{sponge}}$ is added to the original $\frac{\partial A}{\partial t}$, where $A$ is a physical field, $\overline{A}$ is the zonal-mean, and $\tau_\text{sponge}$ is the timescale of the sponge layer. The sponge layer is applied to the three uppermost model levels of three physical fields: zonal velocity, meridional velocity, and potential temperature. $\tau_\text{sponge}$ is 50,000 s, 100,000 s, and 200,000 s for the highest layer, second highest layer, and third highest layer, respectively. 

At the lower boundary, we apply linear Rayleigh drag to represent the drag mechanisms that are believed to limit upper atmospheric wind speeds, such as magnetohydrodynamic (MHD) drag, as discussed in previous literature for exoplanets (\cite{perna2010magnetic}, \cite{menou2012magnetic}, \cite{Menou2012}) and solar system giant planets (\cite{schneider2009formation}, \cite{liu2010mechanisms}, \cite{liu2013predictions}). The Rayleigh drag $- \frac{A}{\tau_\text{drag}}$ is applied to the zonal velocity and meridional velocity of the two deepest layers. The timescale of the bottom drag $\tau_\text{drag}$ is 20 planet days for the deepest layer, and 40 planet days for the second deepest layer, the same as the values in \cite{menou2012magnetic}. We verified through a series of experiments that the conclusions in this paper are not sensitive to the specific values of $\tau_\text{sponge}$ or $\tau_\text{drag}$. Below the deepest layer of the atmosphere, we include a surface layer that has relatively low heat capacity and is in radiative balance with the atmosphere. The heat capacity of this surface layer per unit area is $10^6$~J~K$^{-1}$ m$^{-2}$. For context, this is equivalent to approximately only 0.24 m of well-mixed water on a terrestrial planet such as Earth. The model also includes a convective adjustment scheme \citep{Hourdin2006}. However, as will be discussed in Section~\ref{sec:discussion}, convection is not an important effect in our experiments.  

We initialize the model with wind velocities of zero and an isothermal temperature of 1000 K, which is close to the radiative equilibrium temperature of a 1D gray gas of the deepest atmosphere (1080 K, as can be found by setting $p\to+\infty$ in equation \eqref{eq:graysolution}). We tested different initial temperature profiles, including several isothermal temperatures  (from 500 to 1400 K) and the radiative equilibrium temperature profile of a 1D gray gas, and found similar final results in all cases. We integrated the model for over 250,000 Earth days, to investigate the convergence time. This corresponded to over eight months of computation time on the Harvard Odyssey supercomputing cluster. The key parameters of our simulations are summarized in Table \ref{table:parameters}. The experiments with lower bottom boundary pressure (5 and 10 bar instead of 80 bar) are run for 100,000 days, because they reached equilibrium within a shorter time. We have a set of experiments where the lower boundary drag is turned off. The ``80 bar without drag'' experiment was run for only 150,000 days, because it would have taken us another three months to integrate it to 250,000 days. 
Table~\ref{table:previousModelSummary} summarizes the integration times, radiative schemes, and bottom layer pressure of previous GCM simulations of GJ 1214b. As can be seen, previous studies simulated the atmosphere of GJ 1214b for between 800 and 7800 days only, with assumed bottom layer pressures of between 10 and 200 bar.

\begin{table}[h!]
\begin{center}
\begin{tabular}{c|c}

\hline
\hline
Parameters & Value  \\

\hline

$R_P$ planetary radius (m)  &  $1.7\times 10^7$\\

$g$ gravitational acceleration (m s$^{-2}$)  &  8.93 \\ 

$\Omega_P$ planetary rotation rate (rad s$^{-1}$)  &  $4.615\times 10^{-5}$ \\

$p_{\text{bottom}}$ bottom boundary pressure (bar) & 5, 10, \textbf{80}  \\

\hline

Atmospheric composition  & H$_2$ dominated, $1 \times$ solar \\

$c_p$ specific heat at constant pressure (J kg$^{-1}$ K$^{-1}$) & 13,000\\

$\mu$ mean molecular weight (g mol$^{-1}$) & 2.2\\

$H$ scale height (km)  &  $220$ \\

$\kappa_{\text{sw}}$ shortwave gray opacity coefficient ($\text{m}^2~\text{kg}^{-1}$)  &  $8\times10^{-5}$ \\ 

$\kappa_{\text{lw}}$ longwave gray opacity coefficient ($\text{m}^2~\text{kg}^{-1}$) & $2\times10^{-3}$  \\

\hline 

Horizontal resolution & $64\times 48$  \\

Vertical resolution & $45$ \\

Dynamical time step (s) & 90\\

Radiative time step (s) & 450\\

Total integration time (Earth days)  &  250,000 \\

\hline

\textbf{Note.} Parameters for the default run are shown in bold
\end{tabular}    
\end{center}

\caption{Model Parameters.}
\label{table:parameters}
\end{table}

\begin{table}[h!]

\begin{center}
\begin{tabular}{cccc}

\hline
\hline
Literature & Integration Time& Radiative Scheme & Bottom Layer Pressure \\
 & (Earth days) &  &  (bar) \\
\hline
This paper & 250,000 & Gray gas & 80 \\
\cite{Menou2012} & 7800 & Gray gas & 10 \\
\cite{Kataria2014} & 5000 & Correlated-$k$ & 200 \\
\cite{charnay2015mixing} & 1600 & Correlated-$k$ & 80 \\
\cite{drummond2018effect} & 800 & Correlated-$k$ & 200 \\
\cite{zhang2017effects} & 4000 & Newtonian relaxation & 100 \\
\cite{mayne2019limits} & 1000 & Newtonian relaxation & 200 \\
\hline

\end{tabular}    
\end{center}

\caption{Summary of model setup of previous GJ 1214b 3D simulations of GJ 1214b}
\label{table:previousModelSummary}
\end{table}

% section method (end)

\section{Results} % (fold)
\label{sec:results}

In this section, we describe our model results. As mentioned in the previous section, we performed three sets of experiments, where the bottom layer pressure was 5 bar, 10 bar and 80 bar respectively. We focus on the 80 bar experiment first. 

\begin{figure}[h]
     \centering
     \begin{subfigure}[b]{0.4\textwidth}
         \centering
         \includegraphics[width=\textwidth]{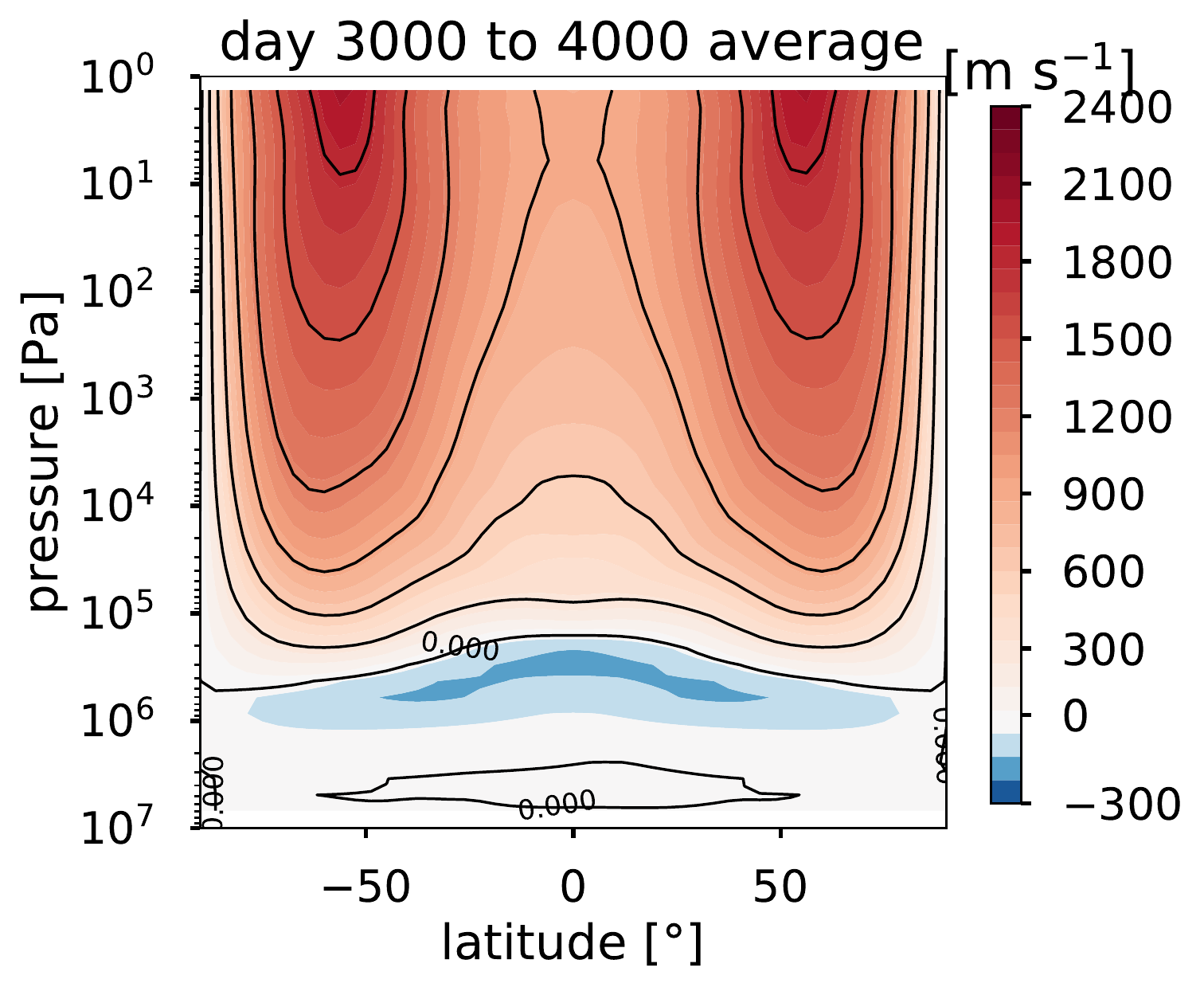}
         \caption{}
         % \caption{our simulation}
         \label{fig:uZonalMeanCompareSubA}
     \end{subfigure}
     \hfill
     \begin{subfigure}[b]{0.4\textwidth}
         \centering
         \includegraphics[width=\textwidth]{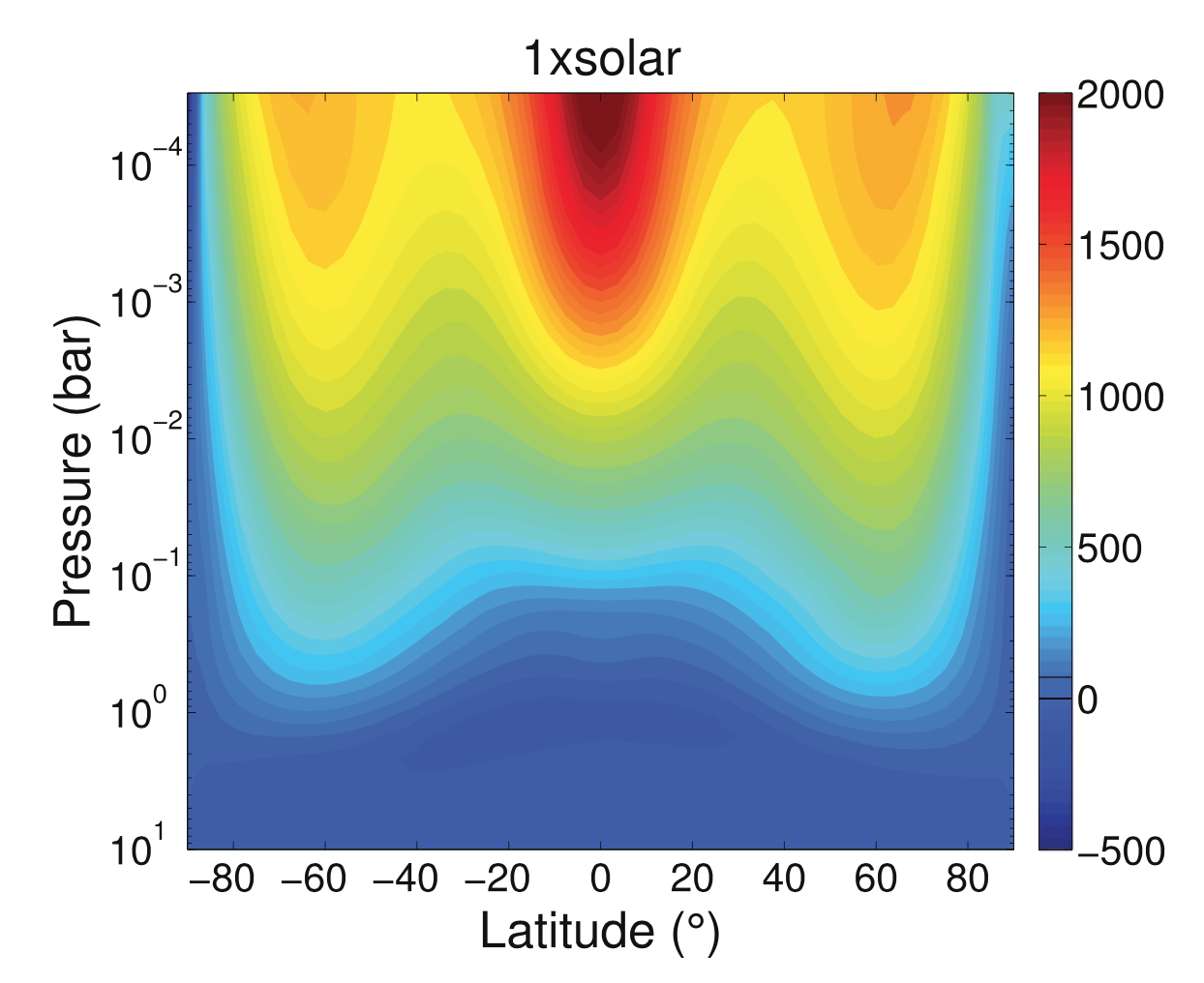}
         \caption{}
         % \caption{Charnay 2015 result}
         \label{fig:uZonalMeanCompareSubB}
     \end{subfigure}
     \hfill
     \begin{subfigure}[b]{0.4\textwidth}
         \centering
         \includegraphics[width=\textwidth]{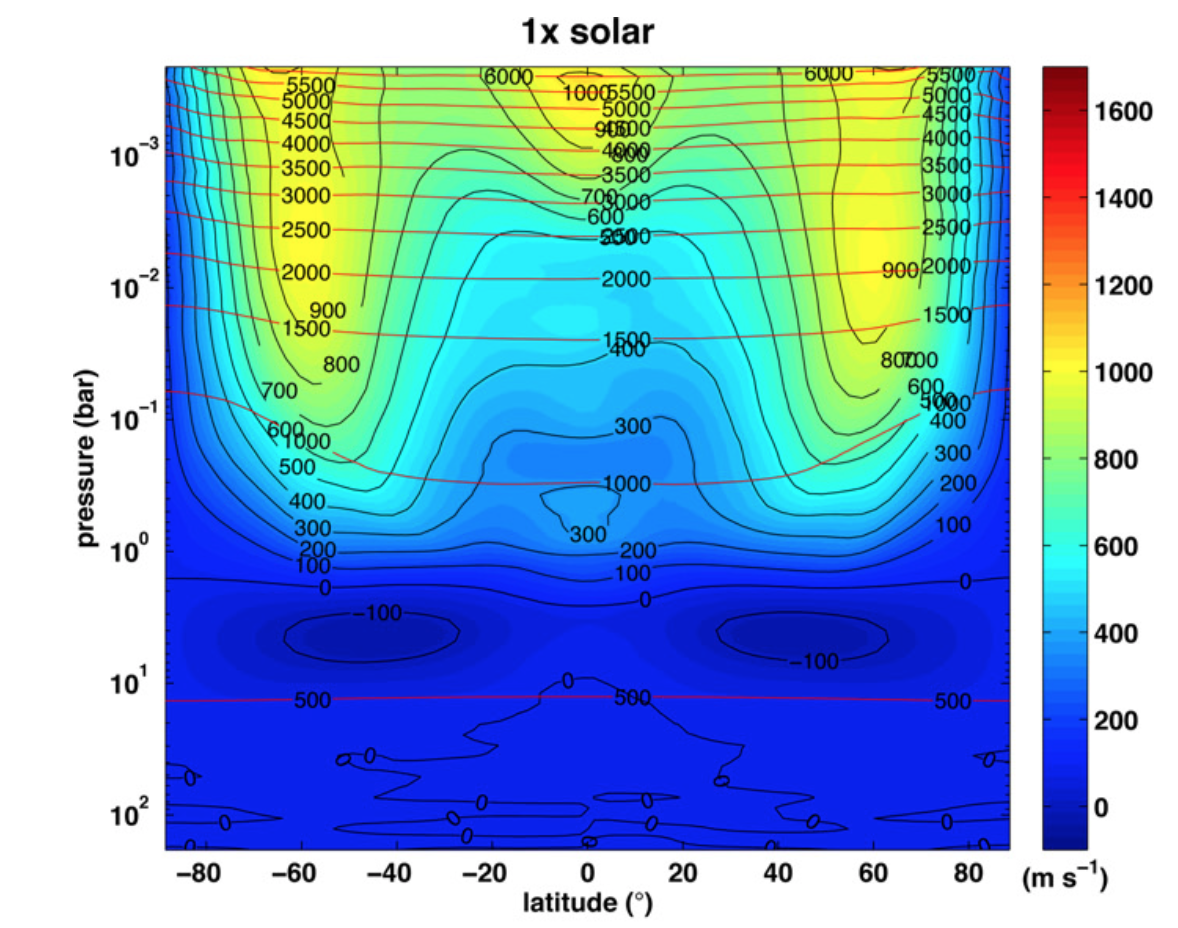}
         \caption{}
         % \caption{Kataria 2014 result}
         \label{fig:uZonalMeanCompareSubC}
     \end{subfigure}
     \hfill
     \begin{subfigure}[b]{0.4\textwidth}
         \centering
         \includegraphics[width=\textwidth]{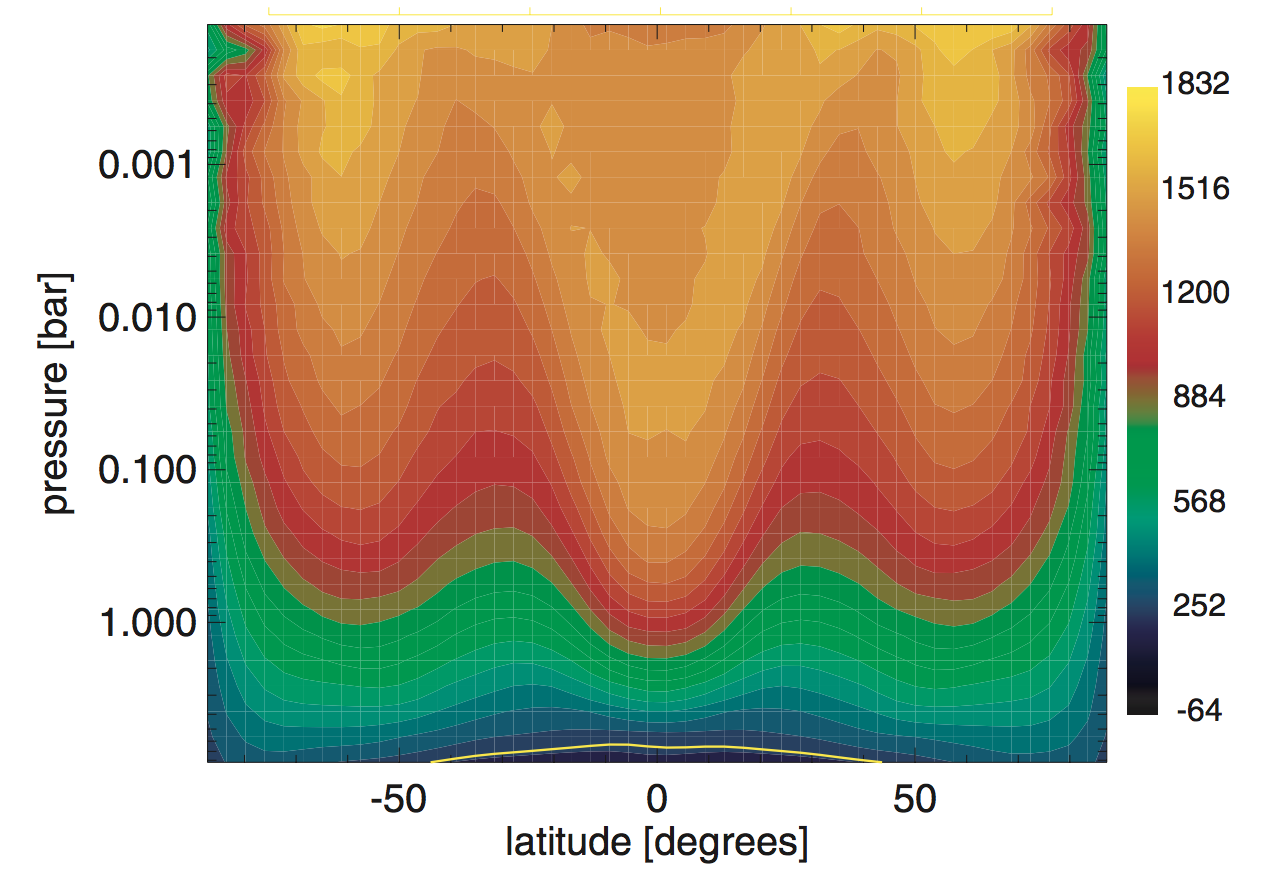}
         \caption{}
         % \caption{Menou 2012 result}
         \label{fig:uZonalMeanCompareSubD}
     \end{subfigure}
     \hfill
     \begin{subfigure}[b]{0.4\textwidth}
         \centering
         \includegraphics[width=\textwidth]{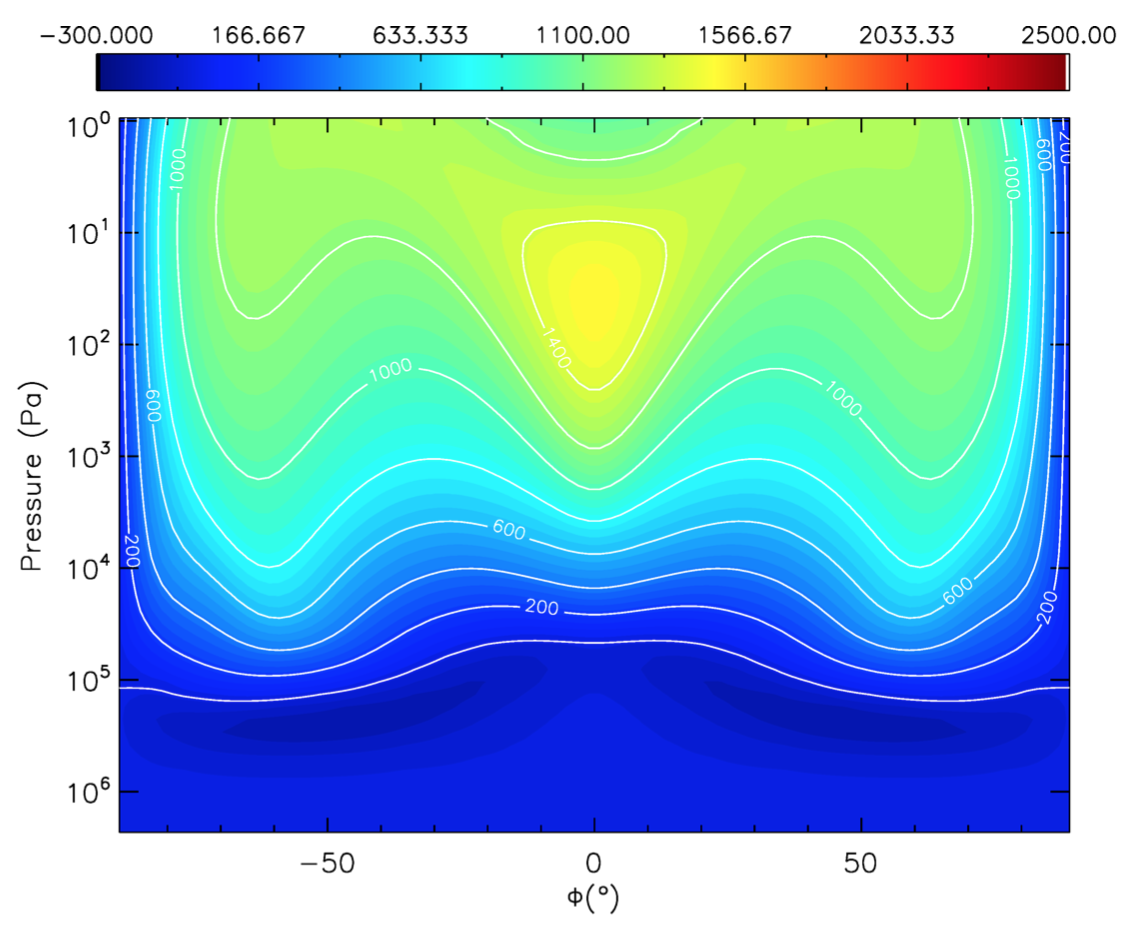}
         \caption{}
         % \caption{Drummond 2018 result}
         \label{fig:uZonalMeanCompareSubE}
     \end{subfigure}
        \caption{Zonal-mean zonal wind for simulations of GJ 1214b with an H$_2$-dominated atmosphere of solar composition, from a range of GCM studies. (a) Our results. (b) Results from \cite{charnay2015mixing}. (c) Results from \cite{Kataria2014}. (d) Results from \cite{Menou2012}. (e) Results from \cite{drummond2018effect}.}
        \label{fig:uZonalMeanCompare}
\end{figure}

On timescales of 1,000-10,000 Earth days, we find comparable atmospheric features to those found in previous studies, especially that they all exhibit two off-equatorial eastward jets. However, over much longer integration time (50,000-250,000 Earth days), we find different atmospheric dynamical features that have significant observational implications. Previous GCM simulations of GJ 1214b with an H$_2$-dominated atmosphere predicted different zonal-mean zonal velocity profiles (Figure~\ref{fig:uZonalMeanCompare}). These previous simulations used different GCM models and made different modeling choices, such as in their radiative schemes, as summarized in Table~\ref{table:previousModelSummary}. Some of the results show a strong equatorial jet, while some are dominated by two off-equatorial jets, depending on the pressure level of interest. 

\begin{figure}[h]
    \centering
    \includegraphics[width=0.8\linewidth]{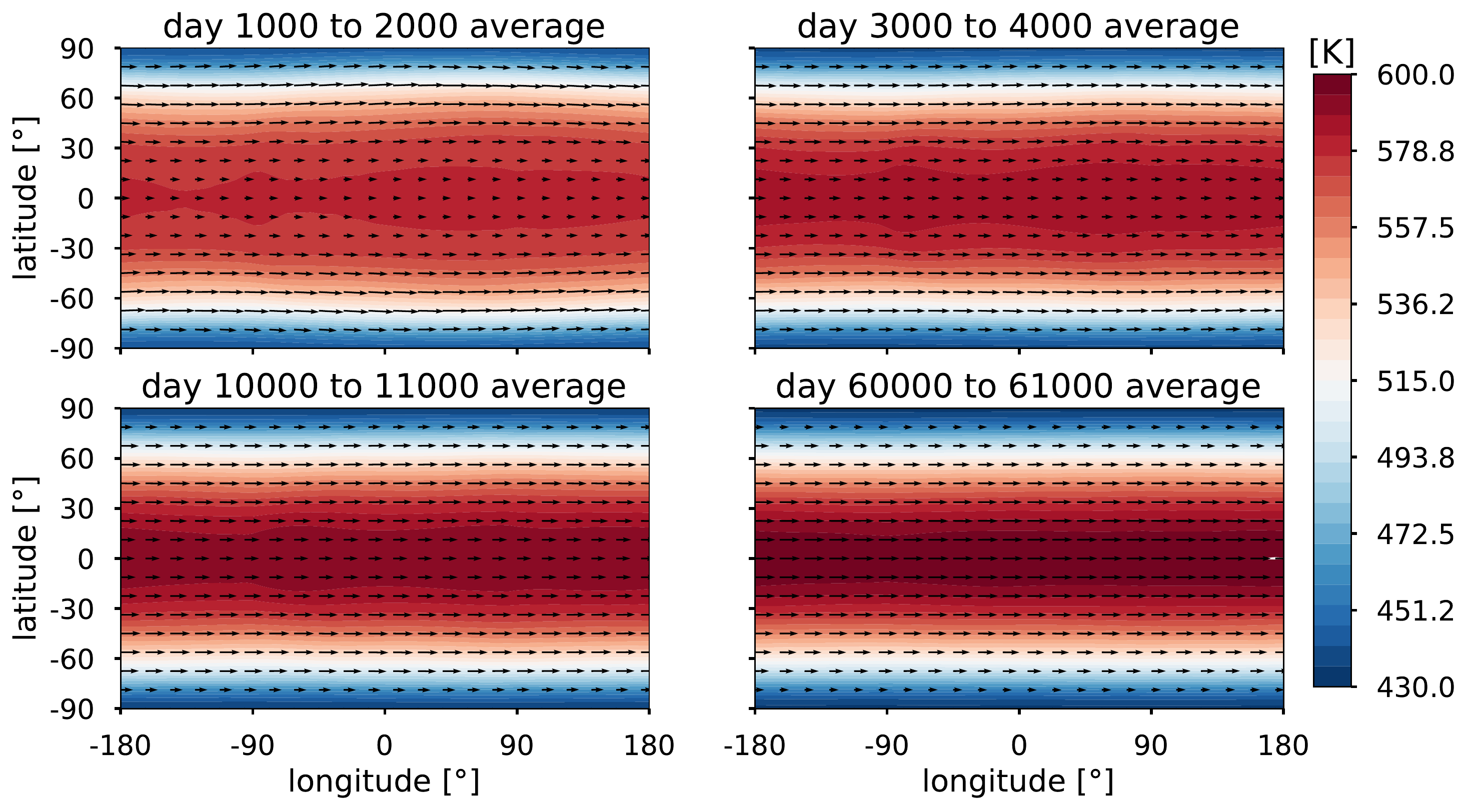}
    \caption{Combined contour-quiver plot of temperature and velocity at 23 mbar, for four different time periods, from our 80 bar GCM simulation. For all four panels, meridional temperature gradients are much greater than longitudinal ones. The isotherms in the upper left panel are visibly less parallel to the $x$-axis than in the other subplots, suggesting greater longitudinal temperature gradients that are later diminished. The velocity quivers in the upper left panel shows two off-equatorial jets, which gradually transition to a wide equatorial jet in the lower right panel.}
    \label{fig:tempLonLat}
\end{figure}

In the upper left panel of Figure~\ref{fig:uZonalMeanCompare}, we show our simulated zonal-mean zonal velocity profile after integrating our model for 3000-4000 Earth days, which is comparable to the typical integration time of previous studies, in the sense that they are all characterized by two off-equatorial jets. The previous results all have equatorial jets in some regions, but they generally do not extend as deep as the off-equatorial jets. The wind and temperature at 23 mbar, as shown in Figure~\ref{fig:tempLonLat}, are very similar to the results of previous studies. The meridional temperature gradient is much greater than the longitudinal one, while the isotherms are not entirely parallel to the x-axis. The wind quivers show two off-equatorial jets. These features are qualitatively identical to the features found in previous studies. As discussed in \cite{Showman2013} and \cite{zhang2017effects}, in the regime appropriate to GJ 1214b, the radiative timescale is much longer than the dynamical timescale (wave propagation timescale). As a result, the day-night temperature contrast is weak and the atmospheric flow self-organizes to form a superrotational zonal jet pattern. 

\begin{figure}[h]
    \centering
    \includegraphics[width=0.9\linewidth]{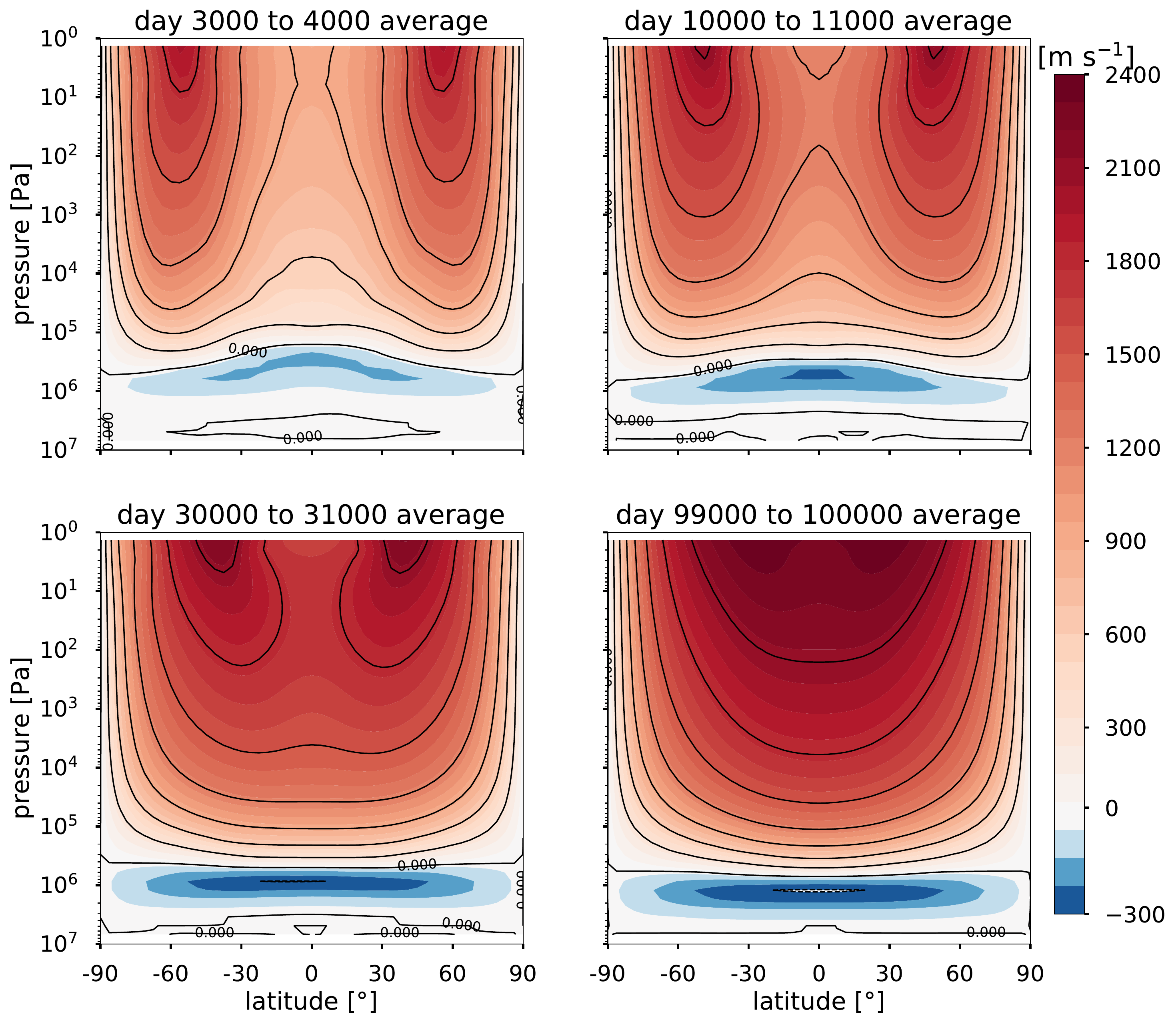}
    \caption{Zonal-mean zonal velocity (m~s$^{-1}$), for four different time periods, from our 80 bar experiment. The plots demonstrate the transition from two off-equatorial jets to one wide equatorial jet. On timescales of 1000-10,000 Earth days, the superrotational jets are centered at around 50$^\circ$-60$^\circ$ latitude. As the system approaches equilibrium, the two off-equatorial jets transitioned into one equatorial jet with higher velocity.}
    \label{fig:uZonalMeanAtTime}
\end{figure}

\begin{figure}[h]
    \centering
    \includegraphics[width=0.9\linewidth]{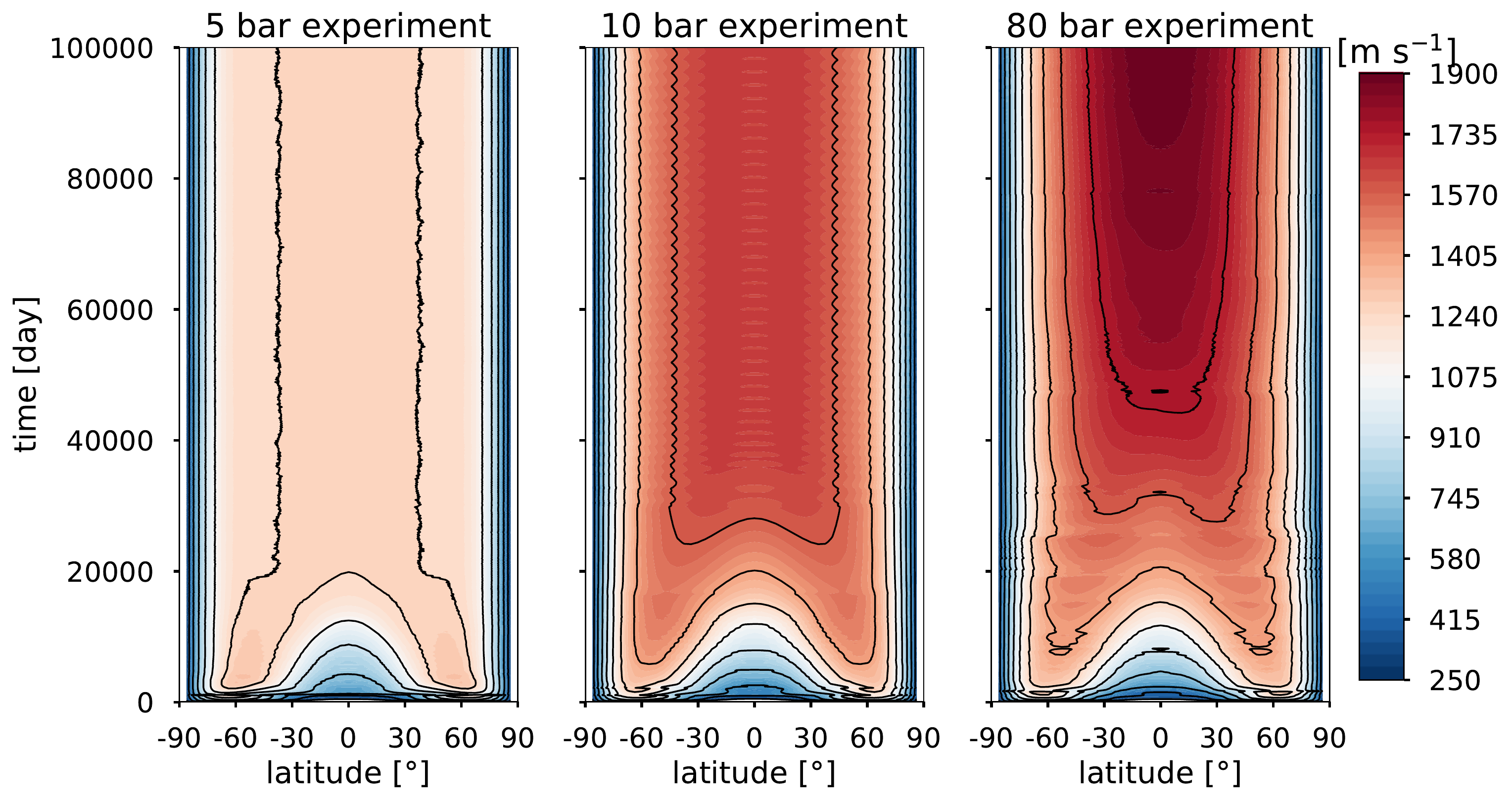}
    \caption{Changes over time in zonal-mean zonal velocity (m~s$^{-1}$) at 23 mbar for three different sets of experiments, with bottom layer pressure of 5, 10 and 80 bar. All three experiments showed transition from two off-equatorial jets to one equatorial jet. In 5 bar and 10 bar experiments, the zonal-mean velocities of this pressure level reached equilibrium after around 20,000-40,000 Earth days. In the 80 bar experiment, the model is still approaching equilibrium after around 100,000 Earth days of integration. The equatorial jet velocity is lower in the 5 bar and 10 bar experiments because the bottom boundary drag was higher in the atmosphere, which effectively lowered the jet speed high in the atmosphere.}
    \label{fig:hovmollerZonalMeanU}
\end{figure}

\begin{figure}[h]
    \centering
    \includegraphics[width=0.6\linewidth]{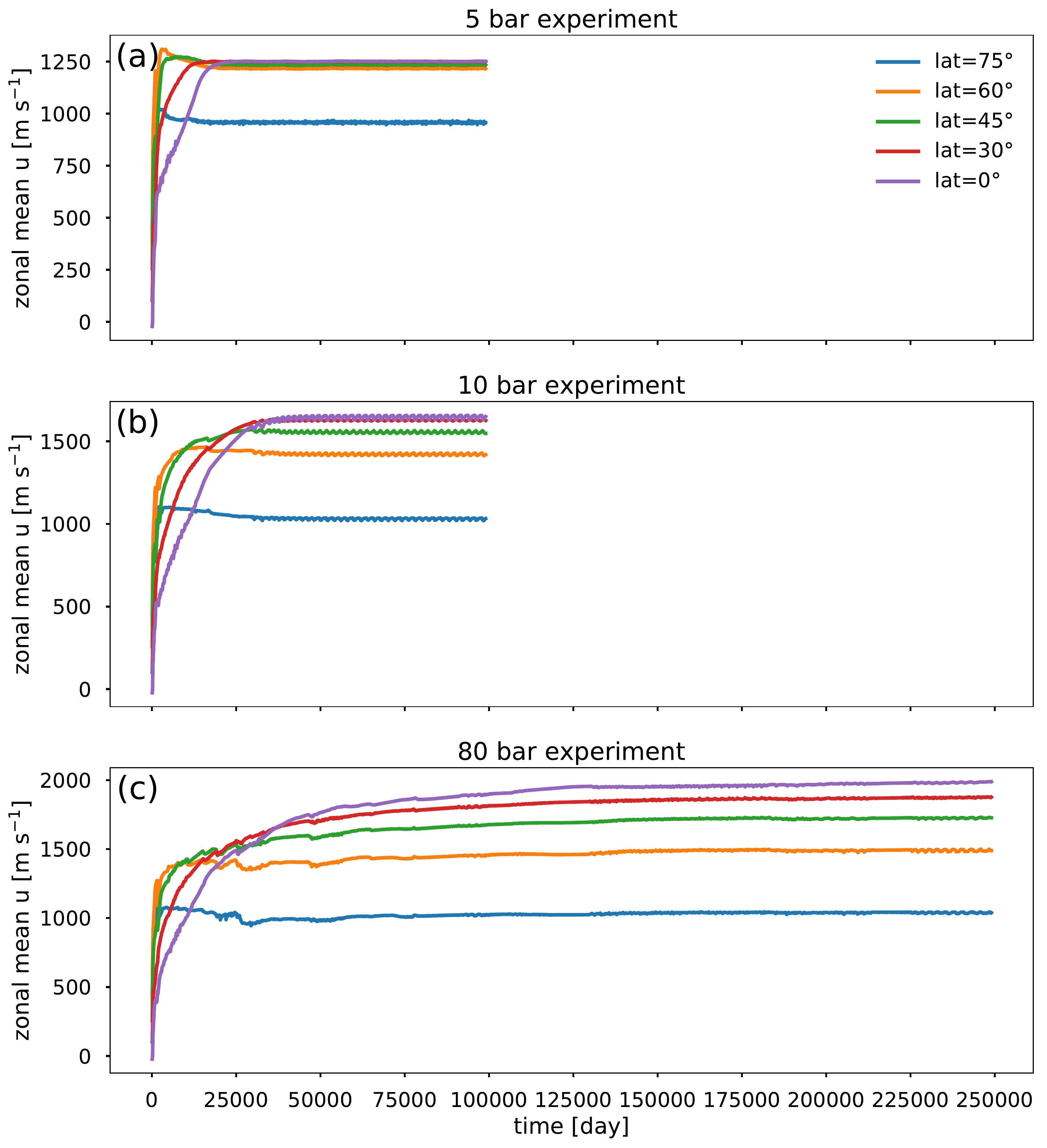}
    \caption{Time series of zonal-mean zonal velocity, at 23 mbar, for three different sets of experiments, with bottom layer pressures of 5, 10 and 80 bar. The plot shows that the zonal velocity at the equator (lat = 0$^\circ$) has a longer equilibrium timescale compared to that at higher latitude (e.g. lat = 60$^\circ$), demonstrating the transition from off-equatorial jets to an equatorial jet. For example, in the 10 bar experiment (middle subplot), the zonal velocity at lat = 60$^\circ$ reaches equilibrium after around 10,000 Earth days. The equatorial zonal velocity (lat = 0$^\circ$) keeps increasing until around 40,000 Earth days. This plot also shows that the 5 and 10 bar experiments require a much shorter time to reach a steady upper-atmosphere velocity. For the 80 bar experiment, the velocities are steady after around 130,000 days.}
    \label{fig:zonalMeanUTimeSeries}
\end{figure}

\begin{figure}[h]
	\centering
	\includegraphics[width=0.6\linewidth]{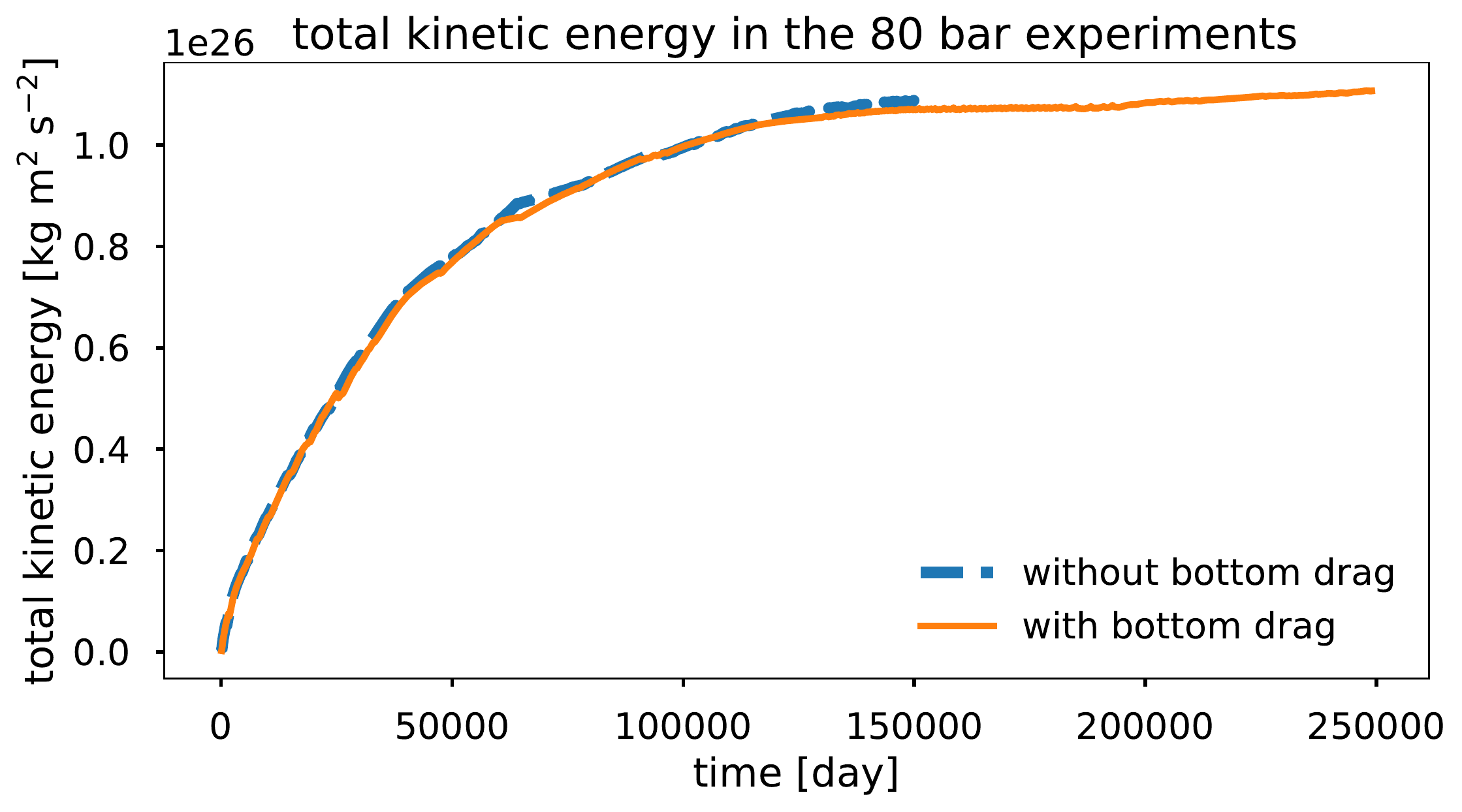}
	\caption{Total kinetic energy of the atmosphere in the 80 bar experiments, showing the long convergence time and the effect of bottom drag. The total kinetic energy steadily increases over time, and then levels off after around 150,000 days. A weak positive trend still exists near the end of the integration time, indicating that the system has not completely converged yet, even though it has been integrated for 250,000 days. The difference between the experiments with bottom drag on and off is very small, suggesting that the effects of bottom drag are not significant within at least the first 150,000 days.}
	\label{fig:kE_80bar}
\end{figure}

After 4000 days, the zonal flow in our simulations was evolving extremely slowly. However, a persistent secular trend was present. We therefore ran the simulation for another 40,000 Earth days. During this long-term integration, we found that a wide super-rotating jet centered at the equator gradually developed, as shown in Figure~\ref{fig:tempLonLat}, Figure~\ref{fig:uZonalMeanAtTime}, Figure~\ref{fig:hovmollerZonalMeanU} and Figure~\ref{fig:zonalMeanUTimeSeries}. This long convergence timescale can also be seen in the plot of total kinetic energy time series in Figure~\ref{fig:kE_80bar}. The continued evolution of the flow features after this long integration time naturally arises from the fact that the deep atmosphere has a long radiative timescale but limited convection. 

\begin{figure}[h]
    \centering
    \includegraphics[width=0.6\linewidth]{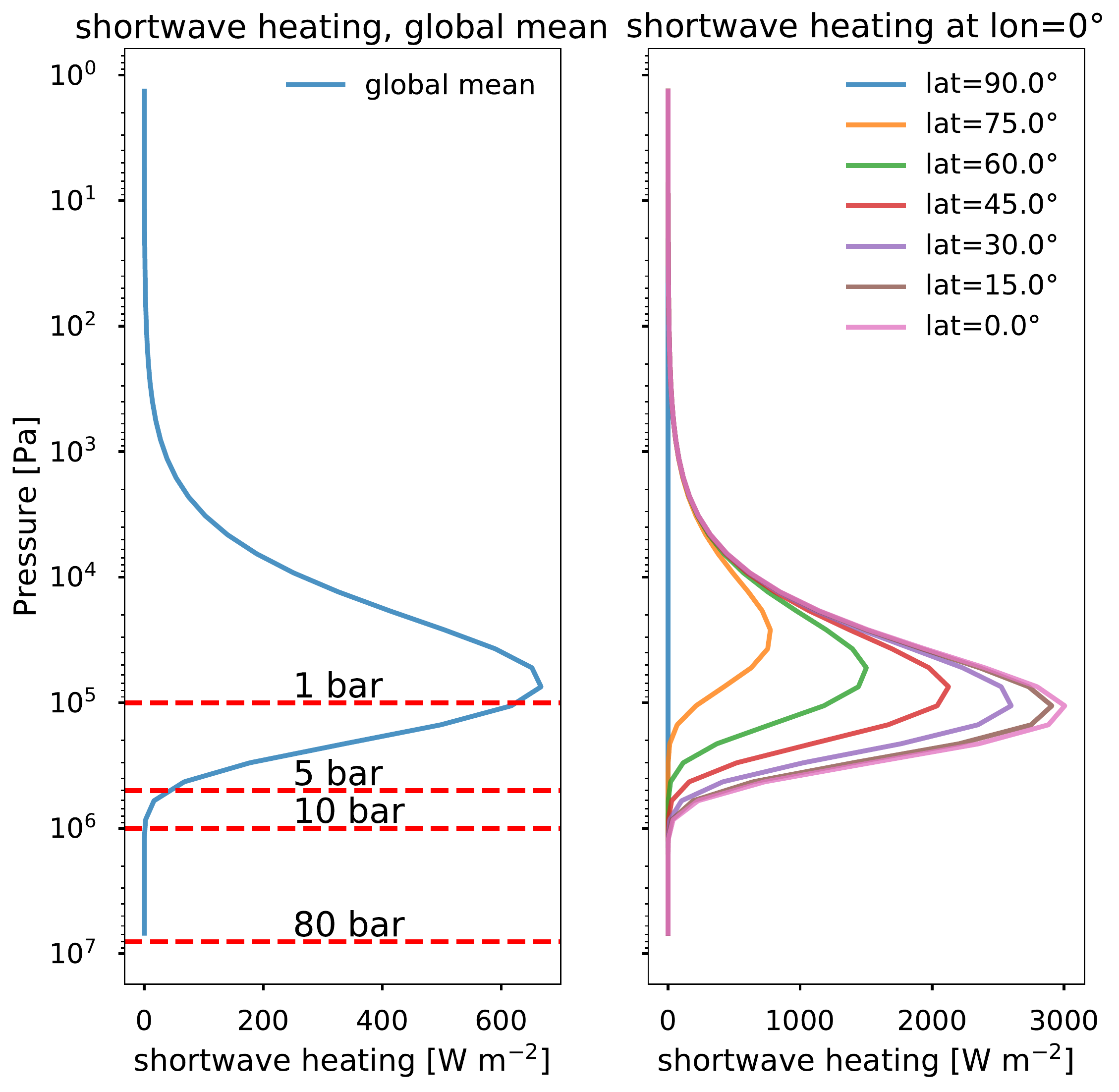}
    \caption{Shortwave absorption profile, for global mean (left panel) and for different latitudes (right panel). The shortwave is mostly absorbed above the 10 bar level, creating a ``stagnant'' region in the deep atmosphere where the equilibrium timescale is very long. The right panel shows that shortwave absorption peaks at different pressure levels for different latitudes.}
    \label{fig:swHeatingVerticalProfile.pdf}
\end{figure}

The radiative equilibrium timescale of the deep atmosphere can in principle be estimated by a scaling analysis, but it also depends on the temperature profile of the rest of the atmosphere, making a first-principles approach challenging. Therefore, we take an empirical approach here. In Figure~\ref{fig:tempChangeRateProfile.pdf}, we use GCM diagnostic data and plot the average absolute temperature change rate $\left|\partial T / \partial t \right|$ due to radiative effects, after integrating the model for 50,000 days. We can see that the rate of change of temperature is very small in the deep atmosphere, due to its high mass and optical depth. From Figure~\ref{fig:swHeatingVerticalProfile.pdf}, we can see that shortwave radiation from the host star is mostly absorbed above the 10 bar level, and the vertical temperature profile deeper than 10 bar level is expected to be very steady. Therefore, the temperature field in the deep atmosphere ($> 10$ bar) is mainly adjusted by radiative effects, which are extremely slow ($< 10^{-8}$~K~s$^{-1}$). We  discuss the roles of convection (which was included in our model) and real-gas effects (which were not) in Section~\ref{sec:discussion}. 

Given the tidally locked forcing pattern, it is natural that the atmosphere develops a meridional temperature gradient with the equator warmer than the poles. Our simulation, like all previous GCM studies, initialized the 3D GCM with horizontally isothermal temperature profiles. Therefore, as the model approaches convergence, a horizontal temperature gradient gradually develops, which takes a long time to form in the deep atmosphere for the reasons discussed above. This meridional gradient of temperature translates into a meridional gradient of geopotential height, which is consistent with the change from two off-equatorial jets to one equatorial jet.

To better illustrate this relationship among the temperature gradient, geopotential height gradient, and zonal jets, we now demonstrate that the flow is in gradient-wind balance by considering the $v$ momentum equation of the primitive equations \citep{Vallis2006}

\begin{equation}
    \frac{D v}{D t} + \frac{u^2}{r}\tan\theta + f u = - \frac{1}{r}\frac{\partial \Phi}{\partial \theta}
\label{eq:mom}
\end{equation}

where $\frac{D}{D t}$ is the material derivative, $u$ and $v$ are zonal and meridional velocities, $r$ is the radius of the planet, $\theta$ is latitude, $f=2\Omega\sin\theta$ is the Coriolis parameter, $\Omega$ is the angular velocity of the planetary rotation, and $\Phi$ is the geopotential height. The geopotential height can be calculated by integrating $\frac{\partial \Phi}{\partial Z} = \frac{R T}{H}$ in the log-pressure vertical coordinate, where $Z$ is height, $H$ is the scale height, and $R$ is the gas constant. 

Since the radiative timescale is much longer than the dynamical timescale in this case, $\frac{D v}{D t}$ is much weaker than the other terms (we confirmed this by checking the simulation results). To make the visualization clearer and symmetric about the equator, we can divide equation \eqref{eq:mom} by $\sin\theta$, yielding

\begin{equation}
    \frac{u^2}{r \cos\theta} + 2 \Omega u = - \frac{1}{r\sin\theta} \frac{\partial \Phi}{\partial \theta}
\label{eq:gradWind}
\end{equation}

\begin{figure}[h]
    \centering
    \includegraphics[width=0.9\linewidth]{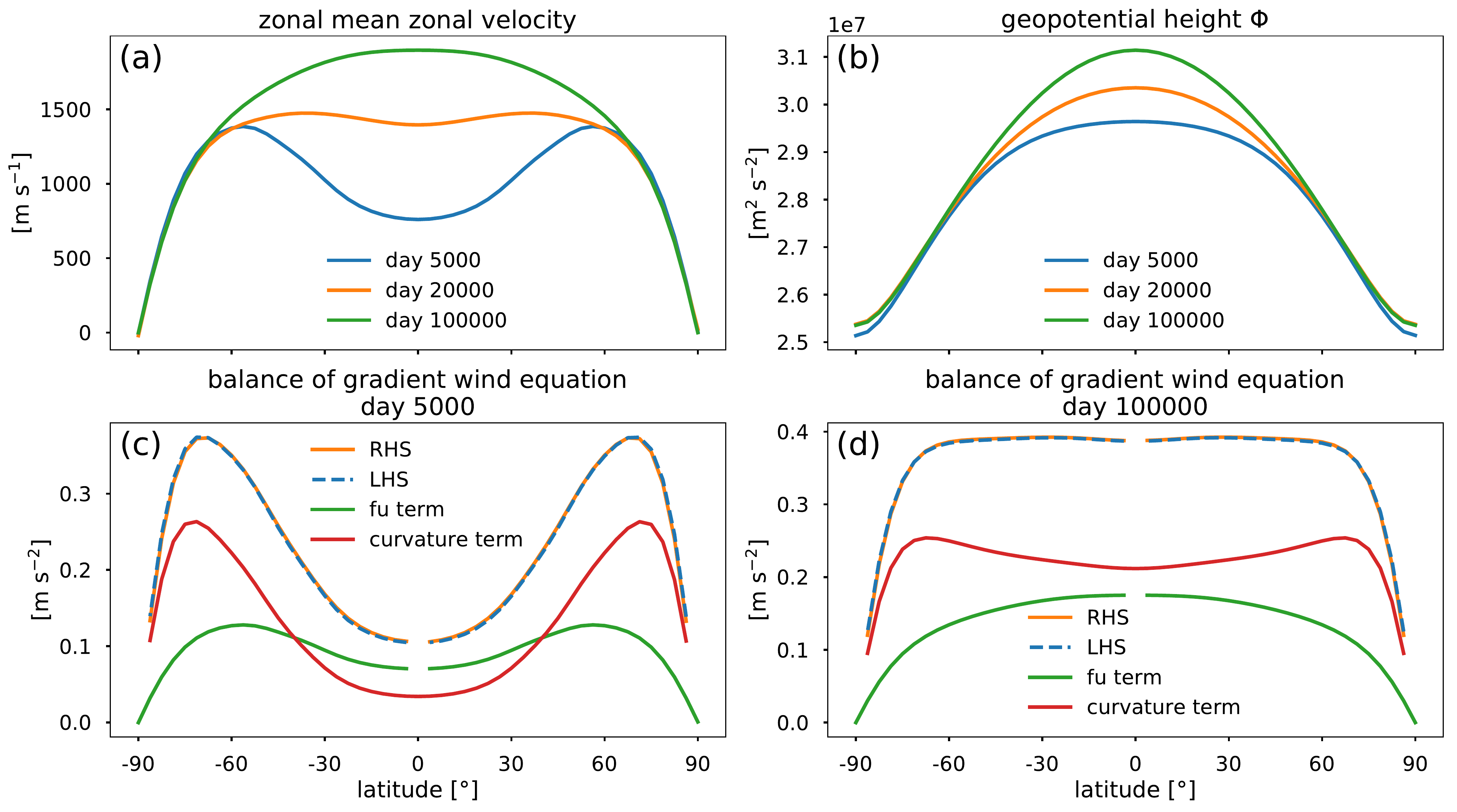}
    \caption{Zonal-mean velocity $u$, geopotential height $\Phi$, and balance of the gradient-wind equation, averaged at 23 mbar. These plots demonstrate that gradient-wind balance holds for both the early and late stages of the experiment. The upper left subplot shows the zonal-mean velocity profile for three different times (day 5000, 20,000 and 100,000), and the transition from two off-equatorial jets to one equatorial jet. The upper right subplot shows the the geopotential height for these three periods. The lower left subplot and lower right subplot show the balance of the gradient-wind equation \eqref{eq:gradWind} for day 5000 and day 100,000 respectively. Early on, in the lower left subplot, the curvature term is larger than the $fu$ term only at high latitudes. Later, in the lower right subplot, the curvature term is larger over all latitudes}
    \label{fig:gradWindBalance}
\end{figure}

In Figure~\ref{fig:gradWindBalance}, we plot the components of equation \eqref{eq:gradWind} alongside geopoential height $\Phi$ and zonal-mean wind $u$ as a function of latitude, for several times in the simulation. As we discussed before, a temperature meridional gradient develops slowly, which in turn increases the meridional gradient of geopotential height at the equator.

\begin{figure}[h]
    \centering
    \includegraphics[width=0.7\linewidth]{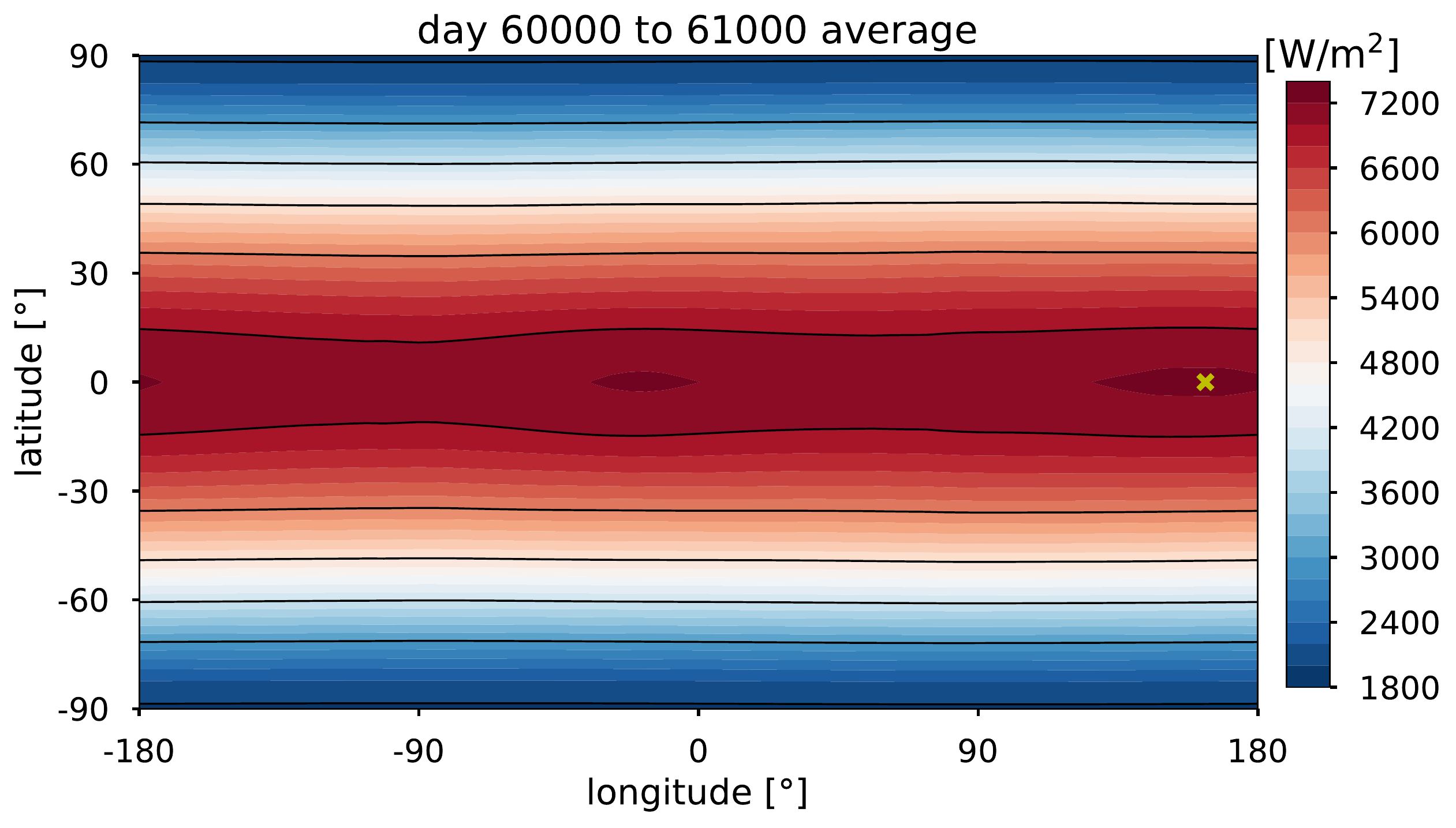}
    \caption{Map of outgoing longwave radiation (OLR). The yellow cross marks the ``hot spot'' at the equator. The meridional variation is much greater than the longitudinal variation, because of the effective transport of heat from the day to the night side by the zonal jets. Note that at the equator, in addition to the hottest spot at around longitude 150$^\circ$, there is another local maximal hot spot at around longitude -20$^\circ$.}
    \label{fig:OLRmap62.pdf}
\end{figure}

\begin{figure}[h]
    \centering
    \includegraphics[width=0.80\linewidth]{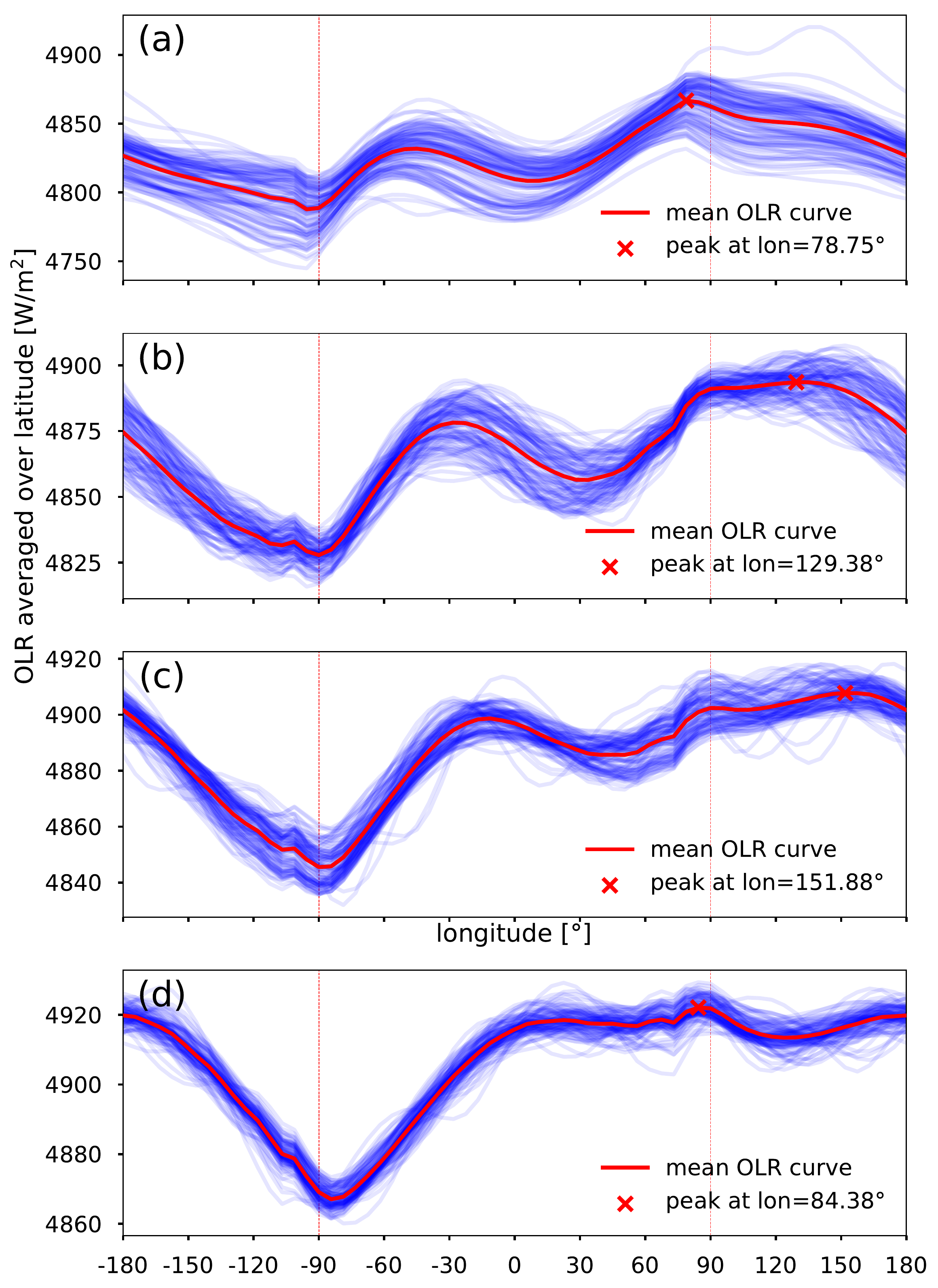}
    \caption{OLR averaged over latitude, after adjusting for the line-of-sight effects. The blue lines are snapshots of every 100 Earth days. The red lines are the mean OLR curve for longer periods (day 10,000-20,000, day 30,000-40,000, day 50,000-60,000, and day 100,000-110,000). The red cross shows the location of the peak of the OLR curve, sometimes called the ``hot spot.'' The vertical dashed red lines indicates the terminator lines.}
    \label{fig:OLRcurveByTimeGroup}
\end{figure}

\begin{figure}[h]
    \centering
    \includegraphics[width=0.60\linewidth]{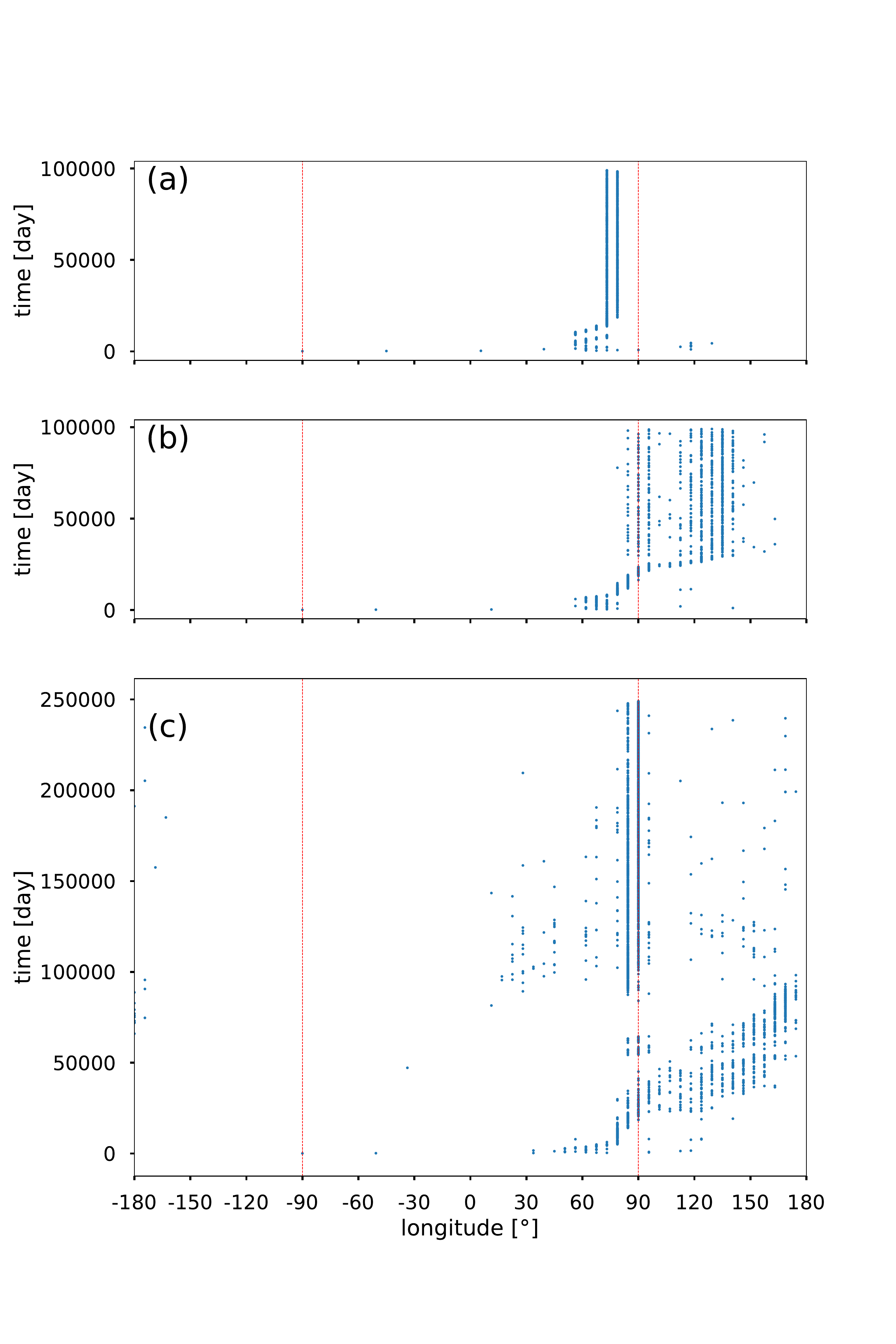}
    \caption{Scatter plot of hot spot location vs. time. In the 5, 10, and 80 bar experiments, we plot the location of the hot spot once every 100 days. The vertical dashed red lines indicates the terminator lines. This plot shows the variability of the hot spot location, and how the hot spot location changes as the model approaches equilibrium. }
    \label{fig:OLRscatterStandardAll.pdf}
\end{figure}

\begin{figure}[h]
    \centering
    \includegraphics[width=0.60\linewidth]{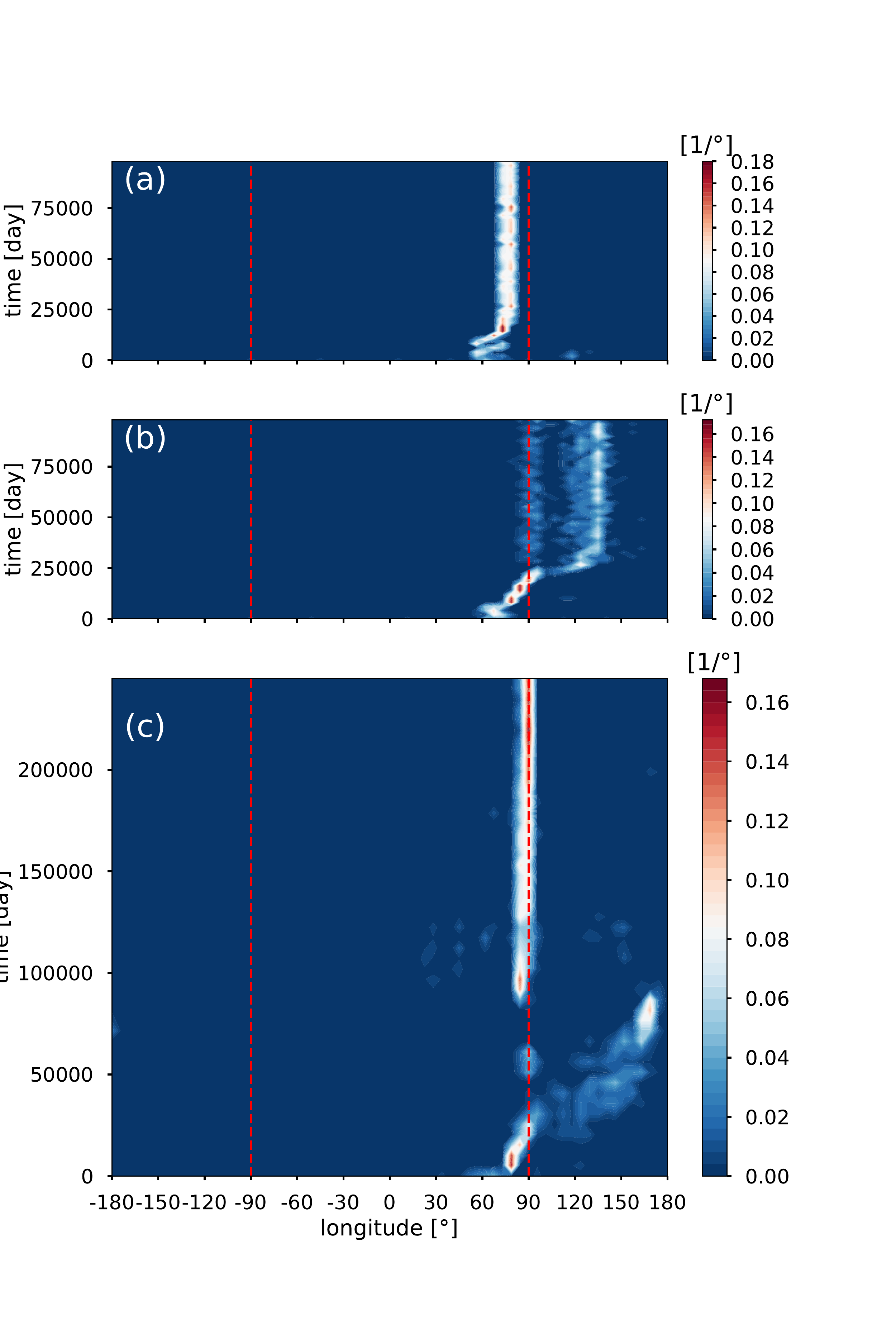}
    \caption{Probability density function (PDF) of the hot spot location for each time period, based on the data shown in Figure~\ref{fig:OLRscatterStandardAll.pdf}. The PDF is calculated for each day and is a function of longitude [deg]. Therefore, it has unit deg$^{-1}$. }
    \label{fig:OLRpdfContour.pdf}
\end{figure}

\begin{figure}[h]
	\centering
	\includegraphics[width=0.6\linewidth]{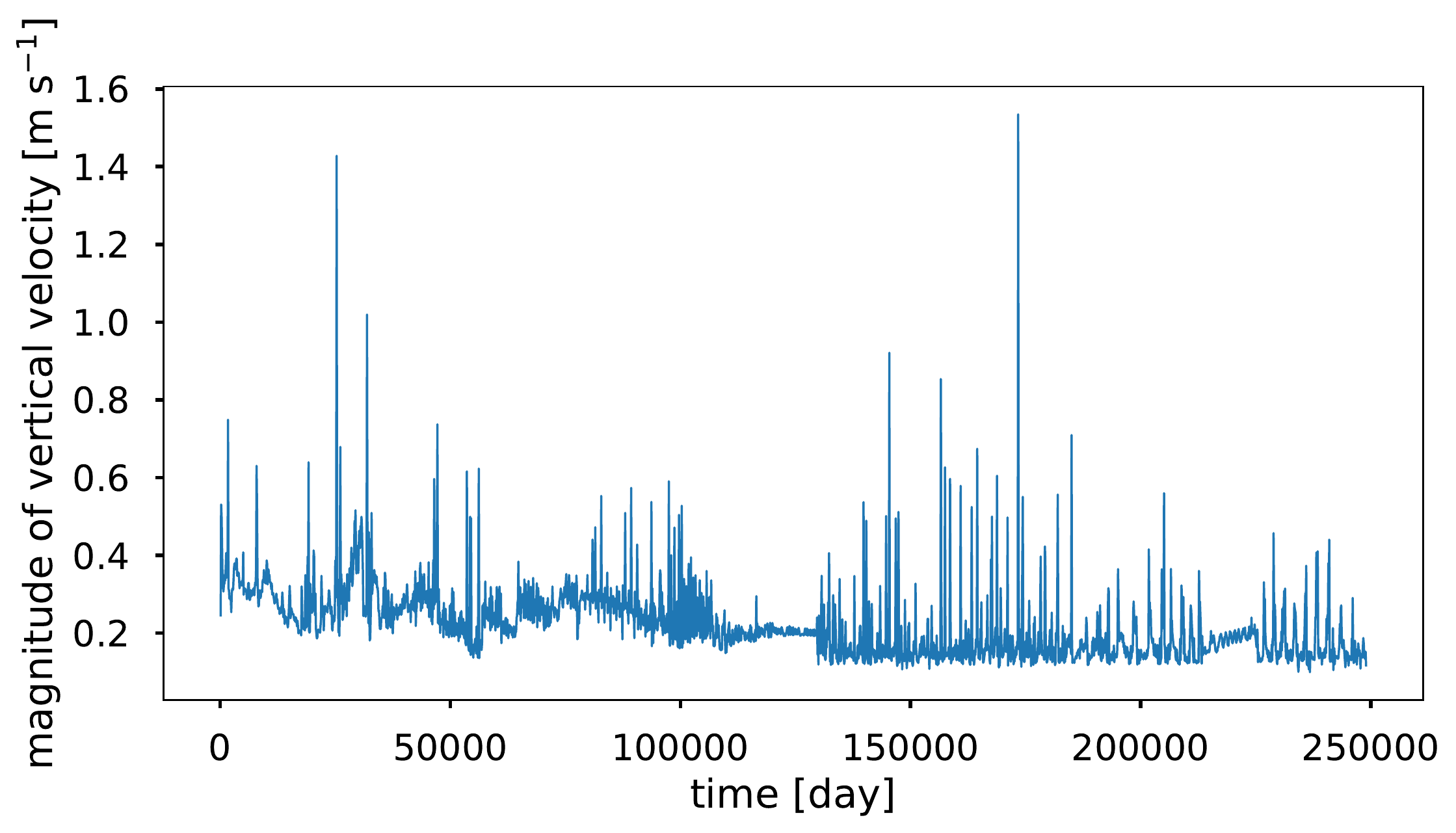}
	\caption{Magnitude of vertical velocity at 23 mbar, showing the strong temporal variability. The time series is sampled every 100 days, which means there are 2500 data points in this plot.}
	\label{fig:w_timeseries.pdf}
\end{figure}

The observational implications of these changes in features are significant. First, as shown in Figure~\ref{fig:zonalMeanUTimeSeries}, the equatorial jet velocity increased from around 1000 m s$^{-1}$ to around 2000 m s$^{-1}$ as we increased the integration time from 10,000 days to 250,000 days. The magnitude of zonal velocity can in principle be directly measured by high-resolution Doppler mapping techniques (i.e. Hot Jupiter HD 209458b \cite{Snellen2010}, HD 189733b \cite{louden2015spatially} and \cite{wyttenbach2015spectrally}). Second, the thermal phase curve becomes flatter because of the increase in zonal velocity over time, as shown in Figure~\ref{fig:OLRmap62.pdf} and Figure~\ref{fig:OLRcurveByTimeGroup}. The locations of the hot spot and the peak in the phase curve also shift eastward between day 25,000 and day 100,000, as plotted in Figure~\ref{fig:OLRscatterStandardAll.pdf} and Figure~\ref{fig:OLRpdfContour.pdf}. This is probably also related to the increased zonal velocity and redistribution of heat. Third, as shown in Figure~\ref{fig:w_timeseries.pdf}, the vertical velocity in the high atmosphere shows strong temporal variability. Given the link between vertical velocity and cloud (e.g., \cite{Charnay2015cloud}), this result suggests that the high cloud coverage could also be intermittent, which might be observable by studying the temporal variability of a transit spectrum. Future studies can quantify the signal-to-noise ratio of these changes and the detectability given certain planetary parameters and observational instruments.

To show the effect of different choices of bottom layer pressure, we ran a set of comparison experiments, where we set the bottom layer pressure to 1, 5, or 10 bar, respectively. We still have 45 vertical levels spaced in log-pressure, and the two deepest levels included the linear drag as described in Section \ref{sec:method}. The 10 bar and 5 bar experiments have qualitatively the same features as the 80 bar standard experiment. In contrast, the 1 bar case does not develop superrotation, and the features are qualitatively very different. Figure~\ref{fig:hovmollerZonalMeanU} and Figure~\ref{fig:zonalMeanUTimeSeries} show that the 10 bar experiment and the 5 bar experiment, similar to the default 80 bar experiment, have two off-equatorial jets initially, which then evolve into a single equatorial jet after a longer integration time. The transition time is shorter for lower bottom layer pressure, which is expected because both the total atmospheric mass and the optical depth have decreased. The superrotation velocity is lower when the bottom layer pressure is lower, because the same bottom drag is applied to the deepest two model layers.

The choice of bottom layer pressure also affects the observables. The wind velocities in the upper atmosphere are very different, as shown in Figure~\ref{fig:hovmollerZonalMeanU} and Figure~\ref{fig:w_timeseries.pdf}. The differences can also be seen in the plots of hot spot location, Figure~\ref{fig:OLRscatterStandardAll.pdf} and Figure~\ref{fig:OLRpdfContour.pdf}. In the 5 bar experiment, the hot spot is always on the day side and has weak variability. In the 10 bar experiment, the hot spot started on the day side in the early stage of the experiment, and moved eastward to the night side around day 10,000-day 30,000. In the steady state, the hot spot location jumps between the terminator lines and around longitude 120$^\circ$. In the 80 bar experiment, the hot spot location continuously shifts eastward as the system approaches equilibrium, but jumps back to the day side after around day 100,000, which can also be seen in Figure~\ref{fig:OLRcurveByTimeGroup}(d). 

In Figure~\ref{fig:uZonalMeanAtTime}, the deep atmosphere (between 1 and 10 bar) shows a westward jet. Similar features are also seen in previous GCM studies of GJ 1214b, such as \cite{Kataria2014}, \cite{charnay2015mixing}, \cite{Menou2012} and \cite{drummond2018effect}. The deep westward jets first appeared as two off-equatorial jets centered at around 40$^\circ$-60$^\circ$, after 100 days of integration, and were similar to the westward jets in \cite{Kataria2014} and \cite{drummond2018effect}. After around 1000 days of integration, the two off-equatorial westward jets combined into one jet centered at the equator, which is similar to the flow patterns in \cite{charnay2015mixing} and \cite{Menou2012}. We found that switching the bottom layer drag or the top of the atmosphere sponge layer on and off did not change the nature of these westward jets. We tested that these westward jets also appear when the model is initialized with different isothermal temperature profiles (such as 600, 800 and 1200 K). When the lower layer pressure is decreased from 80 bar to 10 bar, the westward jet still develops at the same pressure level. In the 10 bar case, since the lower boundary drag is effectively placed higher in the atmosphere, the westward jet is weaker because of the drag. For the 10 bar experiments without lower boundary drag, the westward jet reaches a similar strength as in the 80 bar case. The results of these tests, together with the fact that similar westward jets appear in previous studies using different GCMs, suggests that these deep westward jets are a physical output of the simulations for the given initial conditions and assumed governing equations rather than a peculiar result of a particular model. 

\begin{figure}[h]
	\centering
	\includegraphics[width=0.6\linewidth]{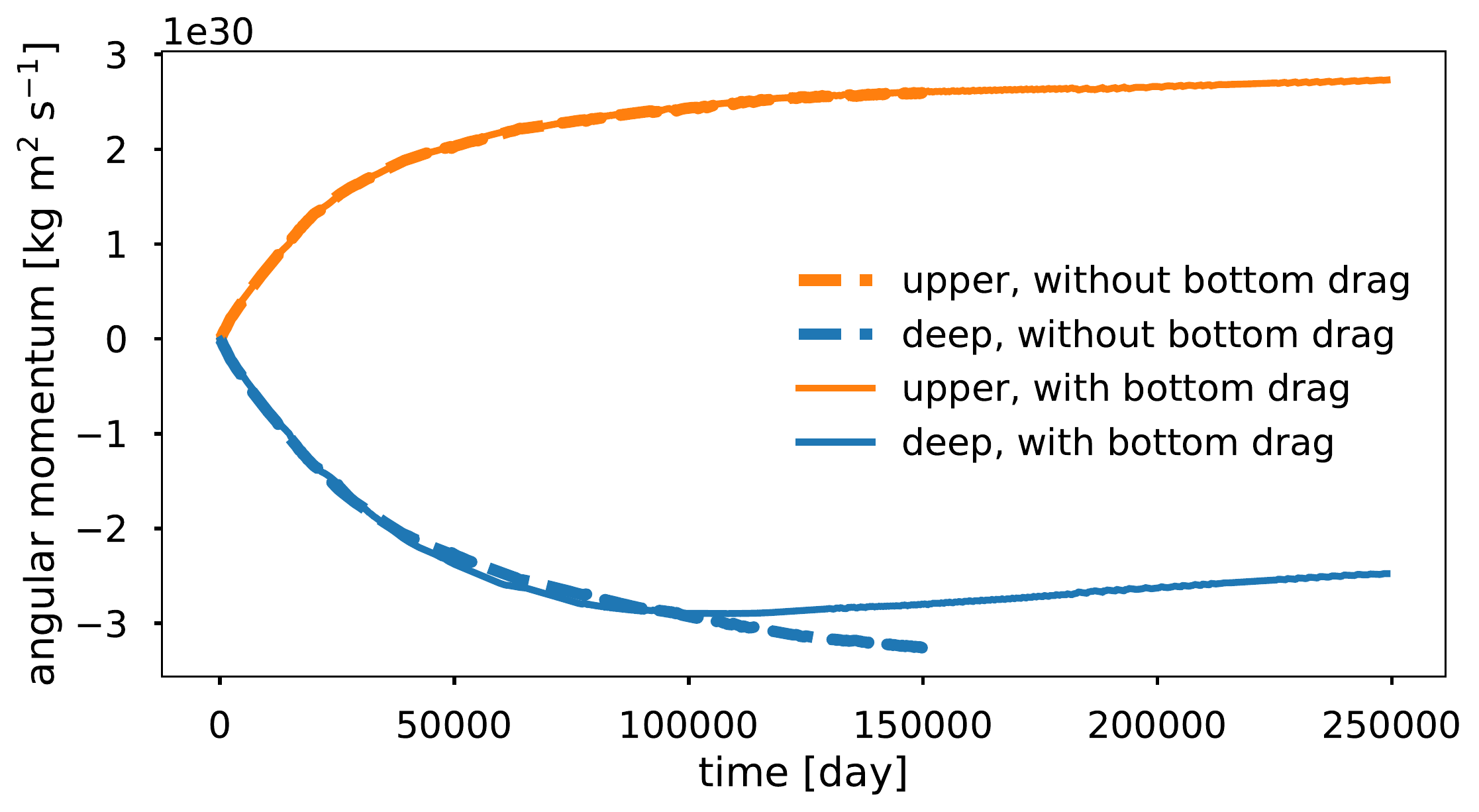}
	\caption{Angular momentum of the upper atmosphere and the deep atmosphere. The pressure separating upper and lower atmosphere is set at 6 bar to most effectively distinguish the eastward flow and the westward flow, as shown in Figure~\ref{fig:uZonalMeanAtTime}. The angular momentum is calculated relative to the surface. This plot shows the eastward acceleration in the upper atmosphere, and the development of the westward jets in the deep atmosphere. This plot also shows the effect of the bottom drag, which is obvious by comparing the angular momentum lines for the deep atmosphere.}
	\label{fig:angMom_80bar}
\end{figure}

Nonethless, deep jet reversals are not seen in the gas giant planets of the solar System \citep{kaspi2018jupiter}. This is probably because on real planets without an artificially imposed boundary at 5-80 bar, negative angular momentum is able to slowly propagate downward until it eventually becomes indistinguishable from the bulk planetary rotation. In a GCM simulation, in contrast, angular momentum exchange can only occur internally or at the bottom boundary. This can be seen by plotting the total angular momentum of the flow in the top and bottom parts of the atmosphere (Figure~\ref{fig:angMom_80bar}). As can be seen, angular momentum of the westward jet in the deep atmosphere nearly balances that of the upper atmosphere. The model conservation of angular momentum is not perfect, but it is close considering the extremely long time-scale of the simulation. We also discovered that switching the bottom layer drag on and off did not change the partition of angular momentum in the first 100,000 days. After that the angular momentum lines of bottom drag on/off experiments deviate from each other but only in the lower branch. Our results suggest that angular momentum diagnostics are important, especially for studies of superrotation, and should be routine in exoplanet GCM studies in the future.

\begin{figure}[h]
    \centering
    \includegraphics[width=0.6\linewidth]{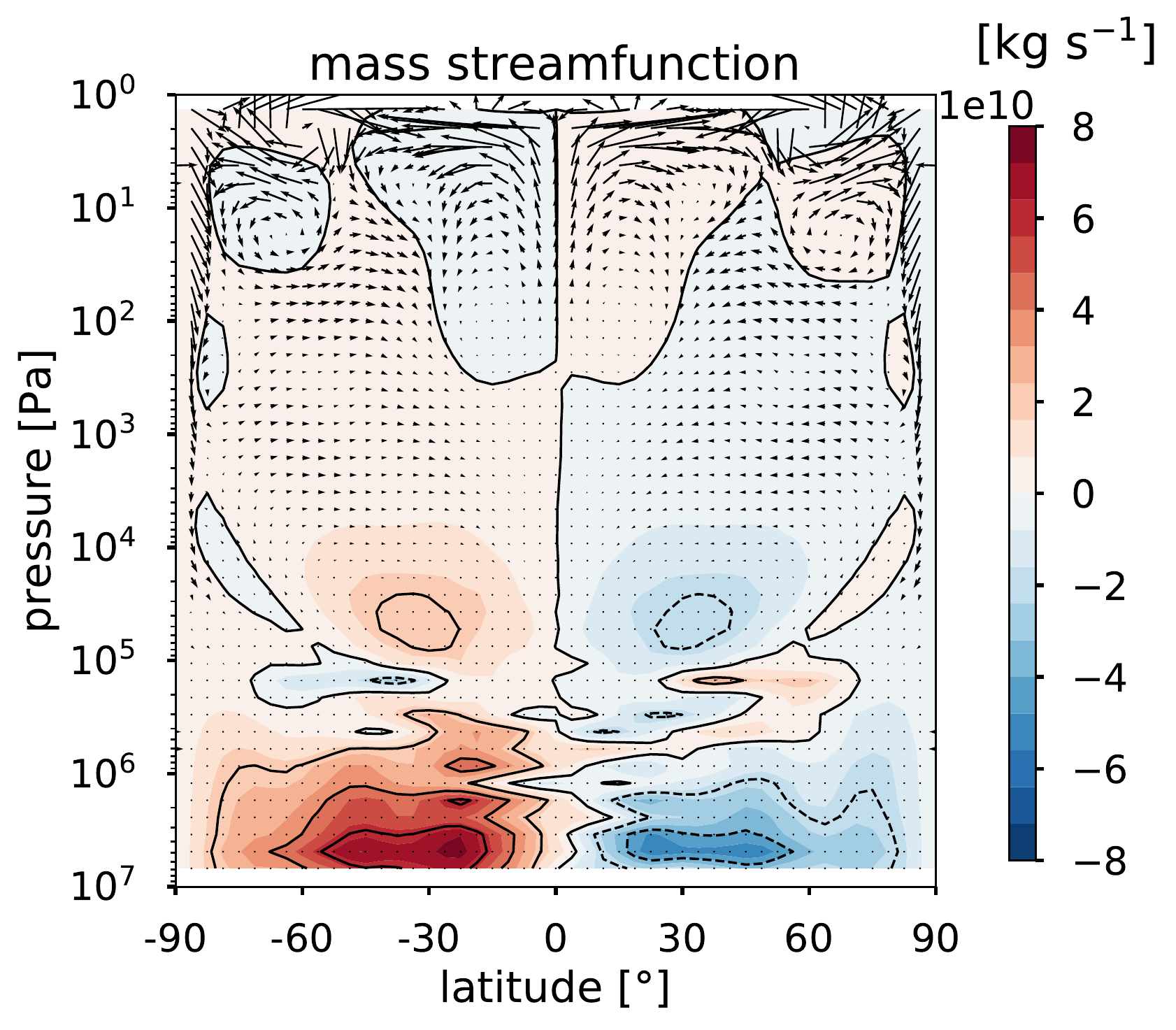}
    \caption{Zonal-mean mass streamfunction. Positive values and solid contour lines correspond to clockwise circulations. Negative values and dashed contour lines correspond to counterclockwise circulations. The quivers are zonal-mean velocities. The circulations have temporal variability, and this plot was calculated with average velocities between day 60,000 and day 100,000.}
    \label{fig:streamfunction60k100k}
\end{figure}

The time-averaged meridional mass streamfunction from 60,000 to 80,000 days is plotted in Figure~\ref{fig:streamfunction60k100k}. In the higher atmosphere, the zonally averaged circulation is characterized by a Hadley-like circulation structure driven by upwelling in substellar regions, and reversed circulation cells at higher latitude. In the lower atmosphere, the circulation cells are downwelling at the equator and the upwelling branch extends to higher latitude. These cells may be mechanically driven by the upper-atmosphere circulation. This trend appears to continue to the deepest atmospheric layer in the model, although accumulation of interpolation error in pressure layer $dp$ and vertical velocity $w$ due to our streamfunction integration may have been significant in this region. We hypothesize that momentum exchange between the upper atmosphere and deep atmosphere mainly occurs via vertically propagating eddy potential vorticity fluxes. A detailed exploration of this issue is left to future work.

\section{Discussion} % (fold)
\label{sec:discussion}

\begin{figure}[h]
    \centering
    \includegraphics[width=0.6\linewidth]{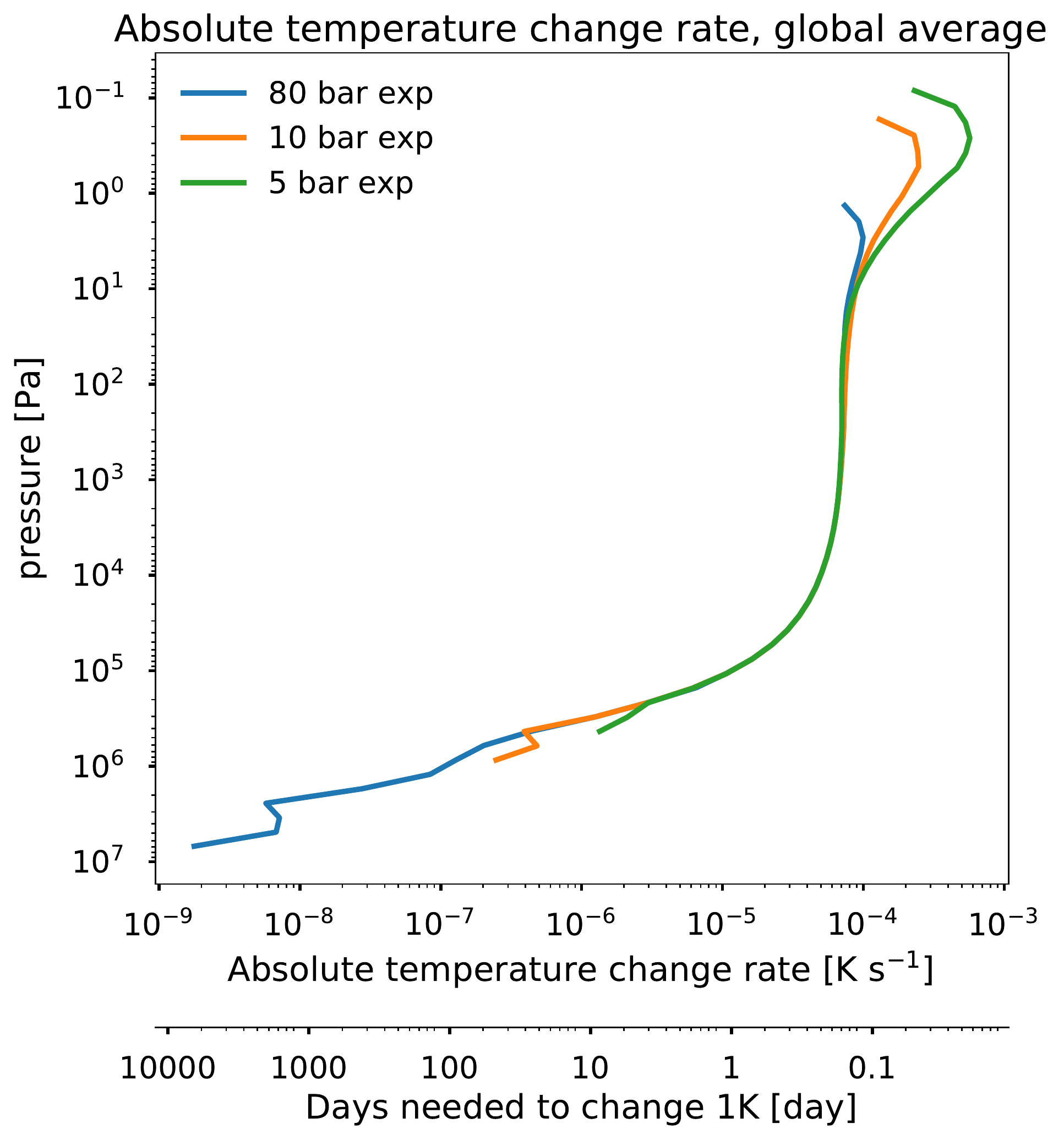}
    \caption{Rate of change of temperature due to radiative effects (longwave and shortwave combined) on different pressure levels, calculated from the GCM after integration for 50,000 days. In the 80 bar experiment, the deep atmosphere requires over 1000 days to change its temperature by 1 K, suggesting that the system has a very long dynamical equilibration timescale}
    \label{fig:tempChangeRateProfile.pdf}
\end{figure}

Our result is based on modeling a hydrogen-dominated atmosphere of GJ 1214b with a gray gas radiative scheme. However, it may have important implications for studies of hot Jupiter planets in general. Most previous GCM studies of hot Jupiters used temperature-pressure profiles from 1D models to initialize their simulations, where the temperature is assumed to be uniform on every isobaric surface. The integration time of the GCM needs to be long enough for the temperature and wind fields to converge to equilibrium. According to the radiative timescale in Figure~\ref{fig:tempChangeRateProfile.pdf}, integrating the GCM for only a few thousand days is equivalent to expecting the difference between the initial temperature field and the 3D equilibrium state to be within a few kelvin. The 1D models cannot provide information on the horizontal temperature structure, and also lack the 3D dynamics in the GCM that can modify the temperature profile. Since the absorbed stellar radiation has a strong horizontal gradient, a horizontal temperature gradient is also expected for most hot Jupiter regimes, where the dynamical timescale is not significantly shorter than the radiative timescale. Deep atmosphere convergence is critical to allowing an accurate prediction of the upper atmosphere and observables. The radiative timescale in Figure~\ref{fig:tempChangeRateProfile.pdf} is calculated based on our simulations of GJ 1214b, which is expected to change with planetary parameters, atmospheric compositions, and radiative schemes. For example, \cite{mendoncca2019angular} found his hot Jupiter simulation converged after 26,500 days of integration, and the hot Jupiter simulations in \cite{mayne2017results} converged in a shorter time. 

We therefore believe that future studies of gas exoplanets will benefit from additional attention to a) model convergence and b) angular momentum exchange between the upper atmosphere and the deep interior. As a minimum, longer integration times and careful monitoring of convergence over longer timescales are recommended. Some approaches can potentially shorten the required integration time, such as a more carefully chosen initial temperature field, or using different integration time steps for different levels of the atmosphere. A simple improvement for the initial condition is to use the temperature profile for 3D radiative equilibrium instead of that for 1D radiative equilibrium. Bottom layer pressure and bottom drag are also important modeling choices that require further study. 

Convection is not an important effect for our experiments, and turning off the convective adjustment scheme does not change the results or equilibrium timescale. The thermal structure of even a strongly irradiated gas planet with an internal flux does eventually transition to a convection-dominated region in the deep atmosphere. There, the temperature profiles are expected to closely follow the convective adiabat, as discussed in many previous papers, e.g. \cite{hubbard1973structure}, \cite{marley2015cool}, and \cite{robinson2012analytic}. According to the radiative-convective equilibrium temperature profile calculated in \cite{miller2010nature}, for an atmosphere of solar composition on GJ 1214b, the radiative-convective boundary (RCB) where the lapse rate of the atmospheric temperature first decreases from an adiabatic rate to a sub-adiabatic value is deeper than 100 bar. We also confirmed using the gray gas opacities of this paper and 1D radiative-convective model that the RCB is deeper that 100 bar. Note that the height of the RCB depends on the intrinsic temperature, which we assume to be 60 K following structure models of \cite{rogers2010three}. This choice matches the energy equilibrium analysis of \cite{thorngren2019intrinsic}, which indicated that the RCB should be much deeper than 100 bar given the planetary parameters of GJ 1214b. Since the bottom layer pressure in our experiments is 80 bar, we do not expect convection to be an important effect in our deep atmosphere.

A difference in the convergence time would be expected if our two-band gray gas scheme were changed to a correlated-$k$ \citep{Showman2009,Wordsworth2011} or even line-by-line radiative scheme \citep{ding2019new}. Some spectral bands might have lower absorption coefficients than others, and these ``window regions'' can change the radiative timescale and the overall convergence timescale. To investigate this possibility, we examined line-by-line absorption coefficients and H$_2$-H$_2$ collision-induced absorption data for a representative atmospheric composition for GJ 1214b (results not shown). We found that unity optical depth in the shortwave occurs around 100 bar for the least opaque wavelength ranges. This is likely an overestimate of the required pressure, because we neglect absorption by important minor species such as H$_2$O. Therefore, even in the real-gas regime, the deeper atmosphere still cannot be effectively heated by shortwave radiation, and will become a convectively stable region with a very long equilibrium timescale. We therefore expect that although the detailed behavior of the jet evolution may vary, the long convergence time we discovered here will still apply to GCM models with correlated-$k$ radiative schemes. 

In this paper, we were modeling a hydrogen-dominated atmosphere, which is an important scenario for many studies of gas exoplanet. For atmospheric bulk compositions other than hydrogen-dominated ones, we must consider the effects of mean molecular weight ($\mu$) and molar heat capacity at constant pressure ($c_p$). Given the same bottom level pressure, a greater $\mu$ means the convergence time is shorter. A higher $c_p$ will result in a longer convergence time. As discussed in \cite{zhang2017effects}, $\mu$ can vary by a factor of $\approx$ 20, while $c_p$ can only vary by a factor of $\approx$ 4. Therefore, if the bottom level pressure and radiative schemes are the same, we expect the hydro-dominated atmosphere to have the longest convergence timescale among all possible atmospheric compositions.

% section discussion (end)

\section{Conclusion} % (fold)
\label{sec:conclusion}

We performed GCM simulations of the exoplanet GJ 1214b and demonstrated that our results resemble previous studies on timescales of 1000-10,000 Earth days, which is the typical integration time of previous GCM studies. Over much longer integration timescales of 50,000 to 250,000 Earth days, we found significantly different flow features. This happened because of the long convergence time in the deep atmosphere, where density is high and radiative timescale is long. The deep atmosphere has a significant impact on the dynamics in the middle and upper atmosphere, as well as on observables such as wind velocity in upper atmosphere, the hot spot location, the thermal phase curve, and potentially cloud coverage.

To properly address the challenge of the long equilibrium timescale of the deep atmosphere for a wider range of exoplanets, further detailed study will be necessary. Conceptually the simplest approach will be to integrate existing models for a longer time and monitor the rate of change of important physical fields, such as temperature and wind velocity. However, it will be important to also consider other model initialization strategies, such as using a 3D radiative equilibrium temperature field, rather than the 1D profile. To reduce computational time, we can also consider reducing the radiative timescale in a physically consistent manner, because the ratio between radiative timescale and dynamical timescale is believed to control important flow patterns such as superrotation (e.g. \cite{zhang2017effects}). Another possibility could be to allow different integration time steps for different layers of the atmosphere, based on, e.g. the approach used in coupled atmosphere-ocean climate modeling \citep{bryan1984accelerating}. Using different time steps for different levels is not yet supported by standard exoplanet GCMs; a simple version of this idea is similar to the approach used in Earth climate GCMs where an atmosphere model is coupled to a dynamic ocean model. Besides, it will be beneficial to test different characteristic timescales of the bottom boundary drag, which might affect the equilibrium timescale in the deep atmosphere. If the bottom boundary drag is strong enough (characteristic timescale is short enough), or applied to higher in the atmosphere, switching on and off the bottom boundary drag could potentially result in very different steady states. Quantitatively budgeting the conservation of kinetic energy and of angular momentum can also help us understand the long-term behavior of the GCM models. For example, \cite{koll2018atmospheric} studied the numerical dissipation of energy and angular momentum, which could guide the future setup of numerical drag and parameterized drag.

As pointed out in Section~\ref{sec:results} regarding the deep westward jet, the interaction between the artificially imposed lower boundary of the atmosphere and the downward propagation of angular momentum to the very deep atmosphere also requires further study. 

This research was supported by NASA Habitable Worlds grant NNX16AR86G and by the Virtual Planetary Laboratory (VPL) program. The GCM simulations in this paper were carried out on the Odyssey cluster supported by the FAS Division of Science, Research Computing Group at Harvard University. The authors would like to thank Feng Ding, Tapio Schneider, Yohai Kaspi, and Wanying Kang for helpful discussions.

% section conclusion (end)

\acknowledgments

\bibliography{manuscript}
\bibliographystyle{plainnat}

\end{document}